\documentclass[sn-mathphys,Numbered]{sn-jnl}% Math and Physical Sciences Reference Style

\usepackage{graphicx}%
\usepackage{multirow}%
\usepackage{amsmath,amssymb,amsfonts}%
\usepackage{amsthm}%
\usepackage{mathrsfs}%
\usepackage[title]{appendix}%
\usepackage{xcolor}%
\usepackage{textcomp}%
\usepackage{manyfoot}%
\usepackage{booktabs}%
\usepackage{algorithm}%
\usepackage{algorithmicx}%
\usepackage{algpseudocode}%
\usepackage{listings}%
\usepackage{xr}

\theoremstyle{thmstyleone}%

\theoremstyle{thmstyletwo}%

\theoremstyle{thmstylethree}%

\begin{document}

\title[Article Title]{Accelerated Screening of Ternary Chalcogenides for High-Performance Optoelectronic Materials}

\author[1]{\fnm{Chen} \sur{Shen}}
\author*[1]{\fnm{Tianshu} \sur{Li}}
\author[1]{\fnm{Yixuan} \sur{Zhang}}
\author[2]{\fnm{Teng} \sur{Long}}
\author[1]{\fnm{Nuno Miguel} \sur{Fortunato}}
\author[2]{\fnm{Fei} \sur{Liang}}
\author[1]{\fnm{Mian} \sur{Dai}}
\author[3]{\fnm{Jiahong} \sur{Shen}}
\author[3]{\fnm{Chris} \sur{Wolverton}}
\author[1]{\fnm{Hongbin} \sur{Zhang}}

\affil[1]{\orgdiv{Institute of Materials Science}, \orgname{Technical University of Darmstadt}, \orgaddress{\street{Otto-Berndt-Str. 3}, \city{Darmstadt}, \postcode{64287}, \state{Hessen}, \country{Germany}}}

\affil[2]{\orgdiv{School of Materials Science and Engineering}, \orgname{Shandong University}, \orgaddress{\street{Jingshi Road 17923}}, \city{Jinan}, \postcode{250061}, \state{Shandong}, \country{China}}

\affil[3]{\orgdiv{Department of Materials Science and Engineering}, \orgname{Northwestern University}, \orgaddress{\street{Clark Str. 633}}, \city{Illinois}, \postcode{60208}, \state{Evanston}, \country{USA}}

\abstract{Chalcogenides, which refer to chalcogen anions, have attracted considerable attention in multiple fields of applications, such as optoelectronics, thermoelectrics, transparent contacts, and thin film transistors. In comparison to oxide counterparts, chalcogenides have demonstrated higher mobility and \textit{p}-type dopability, owing to larger orbital overlaps between metal-X covalent chemical bondings and higher-energy valence bands derived by p-orbitals. Despite the potential of chalcogenides, the number of successfully synthesized compounds remains relatively low compared to oxides, suggesting the presence of numerous unexplored chalcogenides with fascinating physical characteristics. In this study, we implemented a systematic high-throughput screening process combined with first-principles calculations on ternary chalcogenides using 34 crystal structure prototypes. We generated a computational material database containing over 400,000 compounds by exploiting the ion-substitution approach at different atomic sites with elements in the periodic table. The thermodynamic stabilities of the candidates were validated using the chalcogenides included in the Open Quantum Materials Database. Moreover, we trained a model based on Crystal Graph Convolutional Neural Networks to predict the thermodynamic stability of novel materials. Furthermore, we theoretically evaluated the electronic structures of the stable candidates using accurate hybrid functionals. A series of in-depth characteristics, including the carrier effective masses, electronic configuration, and photovoltaic conversion efficiency, was also investigated. Our work provides useful guidance for further experimental research in the synthesis and characterization of such chalcogenides as promising candidates, as well as charting the stability and optoelectronic performance of ternary chalcogenides.}

\maketitle

\section{Introduction}

Optoelectronic materials have attracted significant attention owing to the global energy shortage and environmental concerns suffered from the excessive consumption of fossil fuels.
By absorbing solar energy, which acts as a clean and sustainable energy source, the internal electric current can be produced and delivered based on the photoelectric effect.
The past several decades have witnessed a phenomenal development of the discovery and the performance modification on a series of optoelectronic materials, $e.g.$, silicon, CdTe for solar photovoltaic modules~\cite{scandale2007high,wu2004high}, and III-V semiconductors for emissive layers in light-emitting diodes~\cite{lester1995high, morkoc1994large}.
In addition to the improvement of conventional optoelectronic materials, the continuous advances of the modern industry cannot be separated from the discovery of novel materials with promising physical characteristics as well as favorable functionalities.
Recently, it has been demonstrated that the processes of novel material discovery and the prediction of their properties can be significantly accelerated by using advanced quantum mechanical methods based on density functional theory, high-throughput (HTP) techniques, and material databases represented by the Inorganic Crystal Structure Database (ICSD)~\cite{belsky2002new}, Materials Projects (MPs)~\cite{jain2013commentary}, and Open Quantum Materials Database (OQMD)~\cite{kirklin2015open}.
Compared to the conventional methods using the experimental trial-and-error strategy, this computational screening process can exhibit higher efficiency and more clear structure-property relationships.

Over the years, a variety of HTP studies have been carried out for optoelectronic materials and led to the subsequent chemical synthesis, as well as the device fabrication in the family of perovskites, antiperovskites, half-Heuslers, full-Heuslers, etc.
In 2017, Zhao \textit{et al.} designed a series of lead-free double perovskites by exploiting the strategy of cation-transmutation and highlighted 11 optimal novel candidates with promising optoelectronic properties~\cite{zhao2017design}.
Some of them, $e.g.$, Cs$_{2}$AgInCl$_{6}$ and Cs$_{2}$AgBiBr$_{6}$ have been experimentally verified in recent years~\cite{liu2021lead,locardi2018colloidal,mcclure2016cs2agbix6,zhou2018synthesis}.
Based on the perovskite structural framework, Han \textit{et al.} also proposed two new classes of pnictogen-based quarternary antiperovskites by employing ion type inversion and anion ordering on perovskite lattice sites~\cite{han2021design}.
Five stable compounds represented by Ca$_{6}$N$_{2}$AsSb were identified with favorable properties for solar absorber materials.
He \textit{et al}. designed 99 novel semiconductors that adopted ordered Heusler structure based on 18-electron rule and HTP calculations and identified a series of desired photovoltaic candidates with suitable band gaps, high visible light absorption, and giant dielectric screening~\cite{he2018designing}. 
Kieven \textit{et al}. investigated I-II-V half-Heusler compounds for the buffer layer of solar cells with HTP calculations~\cite{kieven2010ii}.
On the basis of the examination of band gaps and lattice parameters, they proposed a band gap of $>$ 2 eV and a lattice constant of $\sim$5.9 $\AA$  as two principles to avoid absorption losses and lattice mismatch between the buffer layer and the absorber.

Among the aforementioned efforts supported by the HTP working framework, chalcogenides have been extensively studied in various fields.
Recent investigations on chalcogenides have been extended to the fields of thermoelectrics, optoelectronics, topological phases, electronics, and many others~\cite{chen2013classification,wei2020realization,choi2014lateral}. 
In particular, copper-containing chalcogenides (\textit{e.g.}, Cu$_{2}$X (X = S, Se, and Te), CuAgSe)~\cite{qiu2016cu, shi2019chalcogenides} and other chalcogenides with a diamond-like structure ($e.g.$ ZnTe ~\cite{deng2018high}), have been determined with high thermoelectric performance previously. 
In comparison to the oxide counterparts, chalcogenides originated from larger orbital overlaps between covalent metal-X bonds, possess more dispersed valence bands and lower hole effective masses~\cite{kawazoe1997p}.
Moreover, higher-energy valence bands derived by X-p orbitals allow higher \textit{p}-type dopability in chalcogenides.
Consequently, higher $p$-type mobility can be achieved in chalcogenides. For example, the mobility of Cu-based chalcogenides can be up to 20 cm${^2}$ V$^{-1}$ s$^{-1}$, much higher than that of $< 1$ cm${^2}$ V$^{-1}$ s$^{-1}$ for the majority of Cu-based oxides ~\cite{chakroun2021impedance,wei2022yttrium}. 
It suggests that the family of chalcogenide semiconductors can be further explored for promising optoelectronic materials.
However, in spite of the electronic structures, the number of chalcogenides confirmed in the experiment is still lower than that of oxides due to the degradation concerns occurring during the synthesized process, as well as the decomposition tendency in a moisture environment. 
Therefore, exploring novel chalcogenides with satisfied thermodynamic stability and suitable optoelectronic characteristics is demanded.

In this work, we chose 34 ternary chalcogenide prototypes as initial entries to explore novel crystalline semiconductors for potential optoelectronic applications.
In combination with the strategy of cation substitution, we built a material database containing 400,000 candidates.
After a series of screening processes oriented by optoelectronic properties, including thermodynamic stability, band gap, electronic structure, and efficiency, a total number of 52 candidates were identified as promising absorber materials for solar cell devices. 
Our theoretical study provides a guideline for further experimental research focusing on the synthesis and validation of novel optoelectronic materials.

\begin{figure*}
    \centering
    \includegraphics[width=13cm]{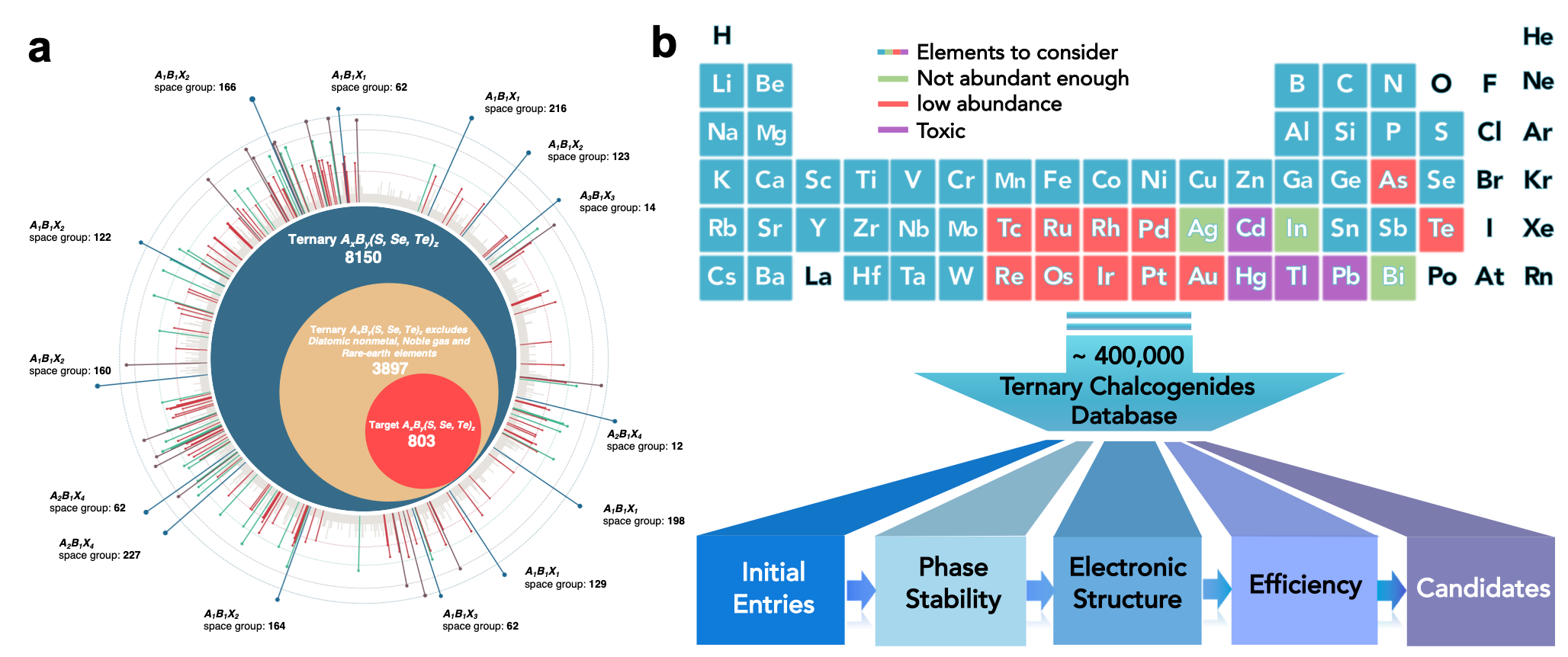}
    \caption{\textbf{a} Crystal structures of ternary chalcogenides in the Material Projects database are classified according to the structural prototypes, and the compounds with a number of more than 30 are highlighted with formulas and space groups. \textbf{b} Theory-driven framework for optoelectronic materials screening. The procedure of constructing ternary chalcogenides by substituting 57 elements in the periodic table, the evaluation of each candidate using the stability and efficiency checks.}
    \label{MP}
\end{figure*}

\section{Results}

\subsection{Crystal Prototypes Selection and Database Generation}

We first collected all the experimentally-known and theoretically-predicted ternary chalcogenide compounds from the MPs database.
As shown in Fig.~\ref{MP}a, there are 8150 ternary chalcogenides existing in the MPs database.
After excluding those cases with the elements of non-metal diatomic molecules (N$_2$ and O$_2$), halogen, noble gas, as well as the rare-earth elements, there are 3897 left compounds crystallizing in 1239 different structural prototypes classified in terms of the chemical formula and space group symmetries. 
Among them, 14 structural prototypes possess significantly more compounds ($>$ 30) than the others ($<$ 5), indicating more promising thermodynamic stability in these 14 prototypes.
Simultaneously, we also considered the synthesizability of ternary chalcogenides and selected 20 well-known ternary phases, which have been verified successfully by experiments. 
The detailed information on these 34 structural prototypes is listed in Table S1.

The generated 34 prototypes with a general chemical formula of A$_{x}$B$_{y}$X$_{z}$ act as initial entries for the HTP material design process.
We then selected the elements from Li to Bi for ion-substitution on A and B sites, respectively, except for the diatomic nonmetal, noble gas, and rare-earth elements.
And one of S, Se, and Te chalcogen anions kept occupying the X site, as shown in Fig.~\ref{MP}.
As a result, we constructed a rough material database consisting of $\sim$400,000 hypothetical candidates for the HTP screening process.

\subsection{Phase Stability Analysis}

Considering that stability is critical for the experimental synthesizability, service lifetime, and efficiency of optoelectronic applications~\cite{bartel2019new,li2017chemically}, we first evaluated the thermodynamic stability of the target chalcogenides.
Previously reported that two criteria can be applied to identify the material that is experimentally promising for chemical synthesis, including the formation energy E$_f$ $\le$ 0 meV/atom and the distance to the convex hull $\Delta E_h$ = 0 meV/atom~\cite{schmidt2017predicting,schleder2019exploring}.
Therefore, as listed in Table~S2, all candidates satisfying these two conditions were screened out, yielding 1417 candidates.
Among them, there are 112, 80, and 33 previously known S-, Se-, and Te-based phases have been successfully validated in experimental, respectively.
From our calculations, 1192 newly-predicted phases with favorable thermodynamic stability were discovered. 
A small subset of these stable ternary chalcogenides was identified in previous computational searches (Ref.~\cite{jain2013commentary}) and has been reconfirmed here.

\begin{figure*}
 \centering
 \includegraphics[height=12cm]{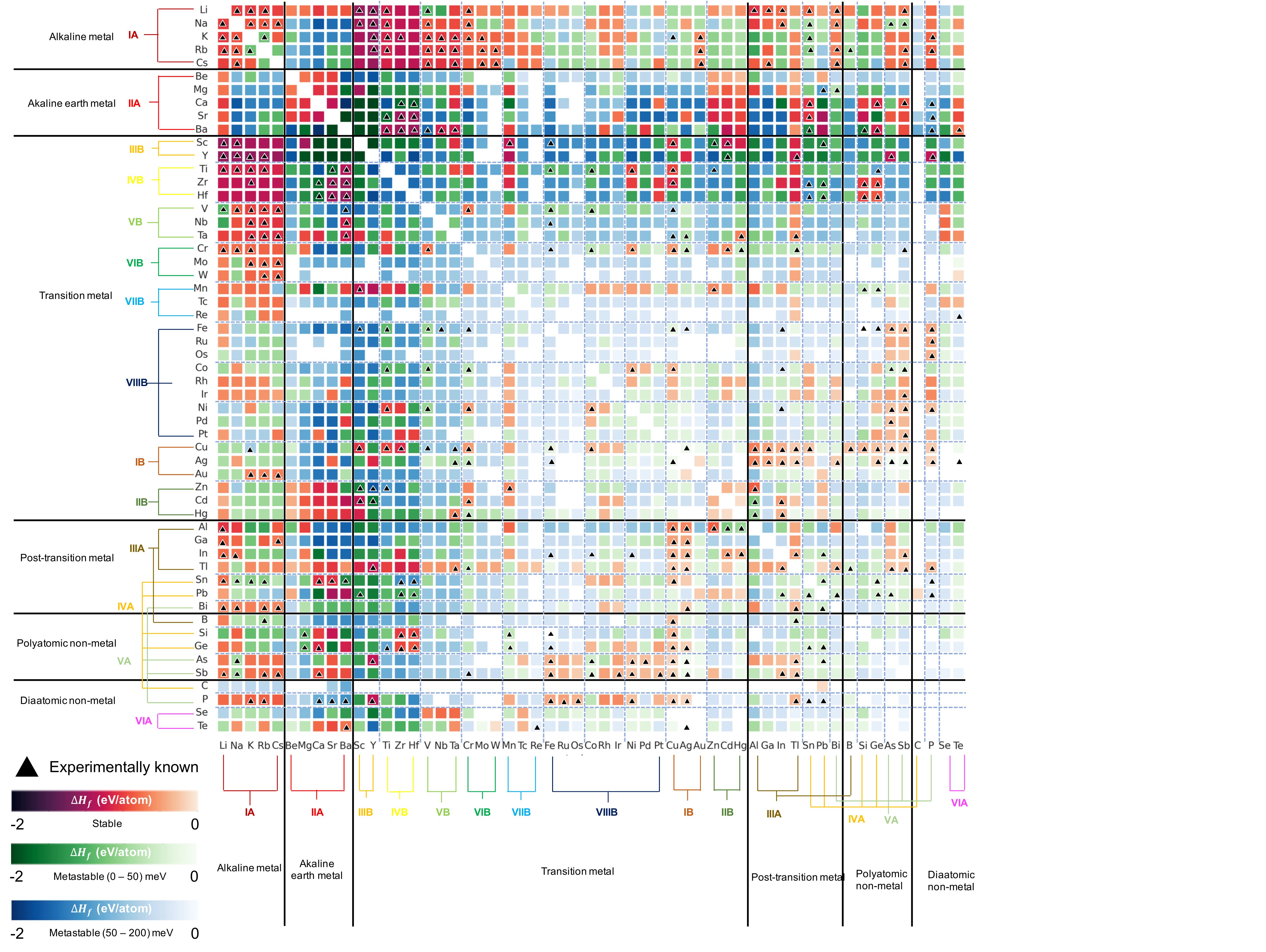}
 \caption{Thermodynamical stability mapping of the S-based ternary chalcogenides. The colors show the lowest formation energy of each candidate, in eV/atom, indicating whether the candidate is stable, metastable, or unstable. The scales are reported in the colorbars at the bottom left. Candidates confirmed by experiments are indicated with black triangles.}
 \label{S-based}
\end{figure*}

Generally, chemical composition and structural symmetry have an in-parallel impact on the thermodynamic behaviors of materials. 
A typical example is the ideally stable halide perovskite with a chemical formula of ABX$_{3}$, which usually exhibits the Goldschmidt tolerance factor ($t=(r_{A} + r_{X})/\sqrt{2}(r_{B} + r_{X})$, where $r_{i}$ is the ionic radii of the atom $i$) close to 1.
This scenario is based on the assumption that the bond lengths of A-X and B-X are the sum of $r_{A}$, $r_{X}$, and $r_{B}$, $r_{X}$, respectively~\cite{sun2017thermodynamic}.
Therefore, A-cation with larger ionic radii ($e.g.$, Cs$^+$ or CH$_3$NH$_3^+$) and undistorted BX$_{6}$ octahedra connected regularly within a 3D structural framework are conducive to the stability improvement of halide perovskites. 
However, it's challenging to intuitively obtain the relationship among the chemical composition, structural symmetry, and physical properties from a long table with detailed results.
Therefore, in the spirit of Pettifor’s phenomenological structure maps~\cite{pettifor1986structures} in conjunction with the work for a map of ternary metal nitrides~\cite{sun2019map}, we generated a multi-functional distance metric considering the thermodynamic stability for ternary chalcogenide candidates with diverse chemical composition.
Then we implemented one-dimensional elemental ordering to construct stability maps of the ternary chalcogenides, colored to represent the stability of the ternary chalcogenides with the lowest formation energy in each A-B-X chemical space.
The red square and the black triangle indicate the stable phase on the convex hull and the experimentally-known phases, respectively.
As shown in Fig.~\ref {S-based} for S-based candidates, it can be found that the stable sulfides locate almost the whole map.
According to the atomic number and groups of elements, the map is distinguished into different blocks (boundaries are highlighted as solid black lines).
In particular, compounds containing alkaline or alkaline-earth elements contribute a considerable amount of the stable phases, especially alkaline and transition metal sulfides represent a majority of the stable spaces.
Whereas stable non-alkali metal–metal sulfides (Me–Me–S) are less common, with small islands of stability scattered amongst the mixed transition- or polyatomic non-metal sulfides.
Notably, the black triangle indicates the experimentally existing stable compounds, suggesting the validity of our methodology and, consequently, the newly-predicted phases.

In addition to the red pixels in the map, metastable ternary sulfides (\textit{e.g.}, green square presents metastable sulfide with a distance to the convex hull below 50 meV, blue square means the metastable sulfide with a distance to the convex hull from 50 to 200 meV), which are able to decompose into competing phases. 
Although metastable phases are generally difficult to synthesize, there are 20 computed metastable phases with $\Delta E_h <$ 50 meV that have been experimentally confirmed, as shown in Fig.~\ref{S-based} by the black triangles in green squares.
After considering a broader convergence region of the convex hull, which is up to 200 meV, metastable candidates are distributed over almost the entire scope.
It should be noted that the critical tolerance with comparable values for the convex hull has also been adopted in other HTP studies~\cite{singh2018high,opahle2013high,wang2017oxysulfide, sun2019map}.
Furthermore, by means of the developed synthesis strategies in experiments, one can expect space expansion for searching targeted  functional sulfide materials beyond equilibrium phases and compositions.
For example, high-temperature strategies using resistive-bearing furnaces~\cite{tian2020high,du2019cobalt,jiang2018isolated} or with s superfast heating speed~\cite{feng2020unconventional} have been extensively utilized for synthesizing metastable nanomaterials, such as singe-atom alloys~\cite{chen2019ruthenium}, high-entropy alloys, and oxides~\cite{yao2020high,yao2018carbothermal}. 
With the ability to control the reaction time, temperature, reactant, and atmosphere, this high-temperature synthesis platform can effectively form metastable compositions while preventing phase separation, coarsening, and ripening~\cite{zheng2022hydrogen}.
Interestingly, the metastable phases mainly fill gaps in the IIIB, IVB, VB, IB, IIB, IIIA, and IVA group-based chalcogenides. 
For Fe, Ru, and Os-based compounds, critical tolerance with considerable values for the convex hull is requested to fill the gap, indicating the demand for a formulated rational strategy for the synthesis. 

Except for the S-based system, we also considered the stability maps for Se-based and Te-based systems, as shown in Figs.~S2-S3.
Compared with the Se-based and Te-based systems, thermodynamically stable ternary sulfides locate on almost the whole map, indicating more promising stability for S-based systems. 
While the stable ternary selenides and tellurides comprise less than a third of the map.
This can be explained by the smaller ionic radius and higher electronegativity of S, resulting in stronger covalent bonding with cations and a consequently more compact crystal lattice framework.
Similarly, alkaline and alkaline-earth compounds contribute a considerable amount of stable phases in both ternary sulfides, selenides, and tellurides. 
While the number of stable transition metal/transition metal chalcogenides (TM-TM-X) is less, scattering amongst the center of the maps for three systems.
Compared with TM-TM-X, the stability of post-transition metal chalcogenides is more promising, and the size of islands of stability is bigger and more concentrated. 
Turning now to the metastable candidates for three systems, metastable sulfides and selenides can fill almost all the gaps in the stability maps. However, the current tolerance of distance to the convex hull still left relative gaps in the chemical composition space of tellurides. For instance, it's hard to find stable or metastable compounds in W/Re/B/C-TM-Te.
Interestingly, there are 2455, 2815, and 2673 compounds for S-, Se- and Te-based chalcogenides exhibiting metastable behavior ( $\Delta E_h < 50$ meV/atom, as shown in Table S1), respectively. More candidates in the predication list and more unoccupied pixels in the tellurides map indicate thermodynamic stable behavior of ternary tellurides has selectivity, which might exist some relationship between chemical composition and valence states, which will be discussed in the next section. 

\begin{figure*}
    \centering
    \includegraphics[height=8 cm]{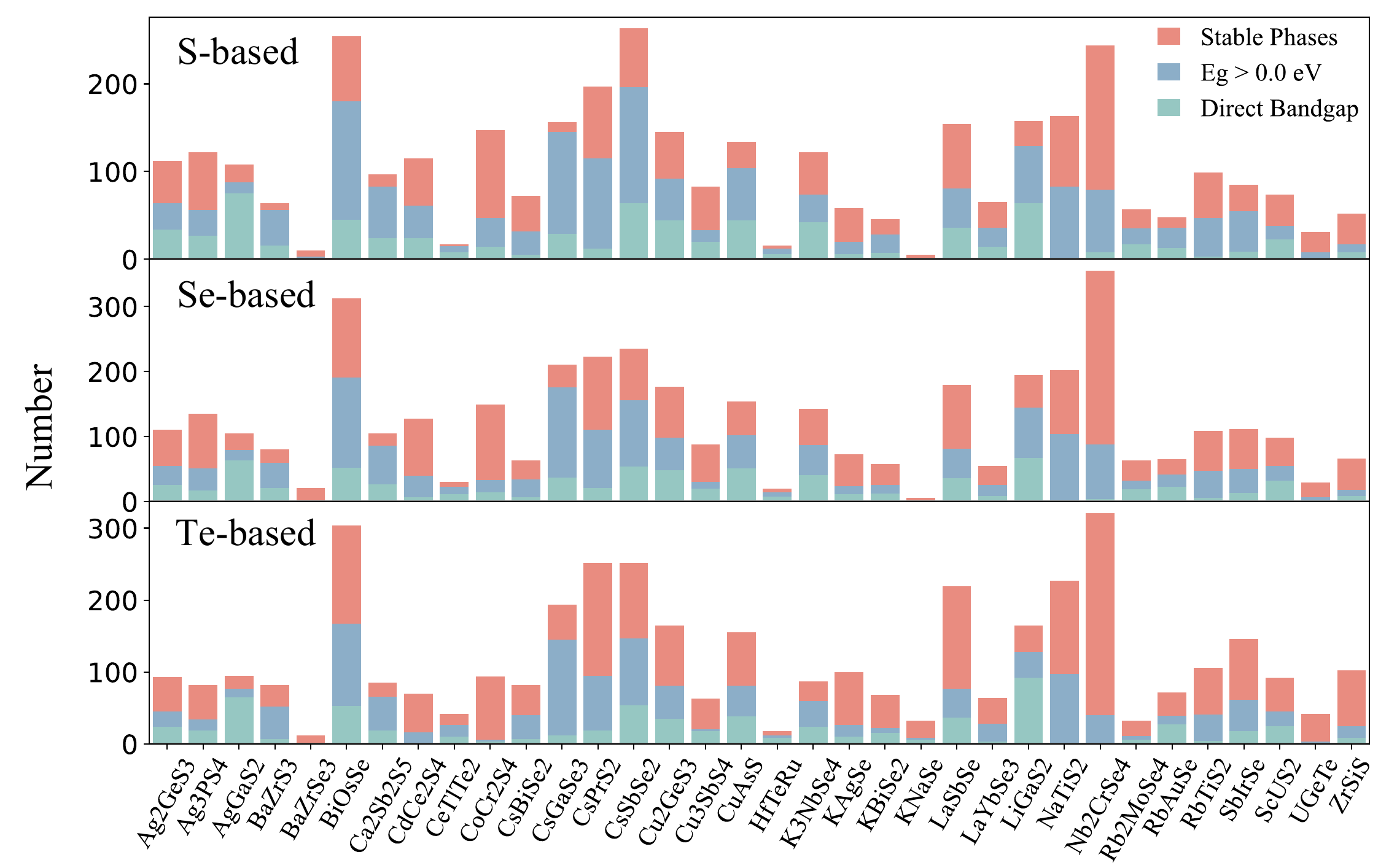}
    \caption{Statistics on different properties of S-, Se-, and Te-based candidates in terms of structural prototypes. Red, blue, and green colors denote the number of stable phases, nonmetallic phases, and direct-bandgap phases, respectively.}
    \label{prototype}
\end{figure*}

Furthermore, we summarized the number of S-, Se- and Te-based candidates with $\Delta E_h < 50$ meV/atom for each structural prototype, respectively, as shown in Fig.~\ref{prototype}a.
As for the relationship between the stability and the structural prototype, BiOsSe, CsSbSe$_{2}$ and Nb$_{2}$CrSe$_{4}$ are three typical prototypes that dominantly contribute to the stable and metastable phases.
Both three structural prototypes exhibit monoclinic systems.
In particular, CsSbSe$_{2}$ and Nb$_{2}$CrSe$_{4}$ are respectively constituted from SbSe$_{3}$ and NbS$_{3}$ tetrahedra, which connect with each other to construct one-dimensional chains.
While the structure of BiOsSe is composed of BiOsSe$_{2}$ tetrahedra interconnected in almost the entire three-dimensional structural space.
Thus, it can be found that the materials containing tetrahedral motifs exhibit a higher potential for stability due to the strong covalent bonding.
The condition of the tetrahedral motifs can also be found in many discovered optoelectronic materials, such as silicon, gallium nitride, and zinc oxide, indicating its validity in promoting structural stability. 
However, it should be noted that the specific arrangement of the tetrahedra and the types of involved atoms are also combined key factors affecting thermodynamical stability, which cannot be determined solely by any of them.

%However, previously reported that most sulfides and selenides are more likely to oxidize and decompose under standard conditions than oxides due to low energy reaction pathways with water or water vapor to form H$_2$S or H$_2$Se~\cite{woods2020wide}.
%A controllable reaction condition, which may involve high temperatures or pressure, is required for the synthesis process of sulfides.

\subsection{Formation Energy Prediction via Machine Learning}
To understand the trend of stability for calculated ternary chalcogenides, we here take the ternary sulfides as an example.
The stability tendency with respect to the chemical composition is analyzed based on the Hume-Rothery rules~\cite{mizutani2012hume} and Miedema’s theory~\cite{bennett1979theory, miedema1980cohesion, bakker1998enthalpies}. 
Such rules and semi-empirical theory are formulated based on the difference of atomic size, electronegativity factors (or \textit{i.e.} the electronegativity difference), the valence electron concentration (VEC), and the electron-density discontinuity. 
Different combinations of such factors are tested, and it can be noticed that there are no consistent stability trends of ternary sulfides along these factors. 
This is because of the diversity of both involved structural prototypes and the chemical compositions, which will cause the failure in exploring the impact of stability factors.
After this, we took the prototype with a chemical formula of A$_{1}$B$_{1}$X$_{1}$ act as an example for checking the valence states influence by plotting the Pettifor’s structure map, as shown in Fig.~S4. As one can see, it's hard to determine the distribution of stable candidates. And, we only could define a forbidden zone for compound formation, namely TM-TM-X.
Therefore, the selection of a general model plays a key factor in solving this problem for the current database.
In this regard, a stack-ensembling ML strategy based on the AutoGluon framework~\cite{erickson2020autogluon} is implemented in our work to explore the decisive descriptors affecting the thermodynamic stability of ternary chalcogenides.
Compared to a single deep neural network troubled with prone overfitting concerns, this automatic stack ensembling framework can maximize the accuracy of prediction, and the ensemble features can mostly guarantee the generalizability of the model, leading to the high accuracy of the prediction.
Therefore, we developed an ML model to accurately predict the E$_f$ of chalcogenides using our generated database.
The trained model was further applied to facilitate the rapid screening and prediction of unexplored chalcogenide structures, thereby providing valuable assistance in the discovery of novel materials. 

\begin{figure*}
    \centering
    \includegraphics[height=4.0 cm]{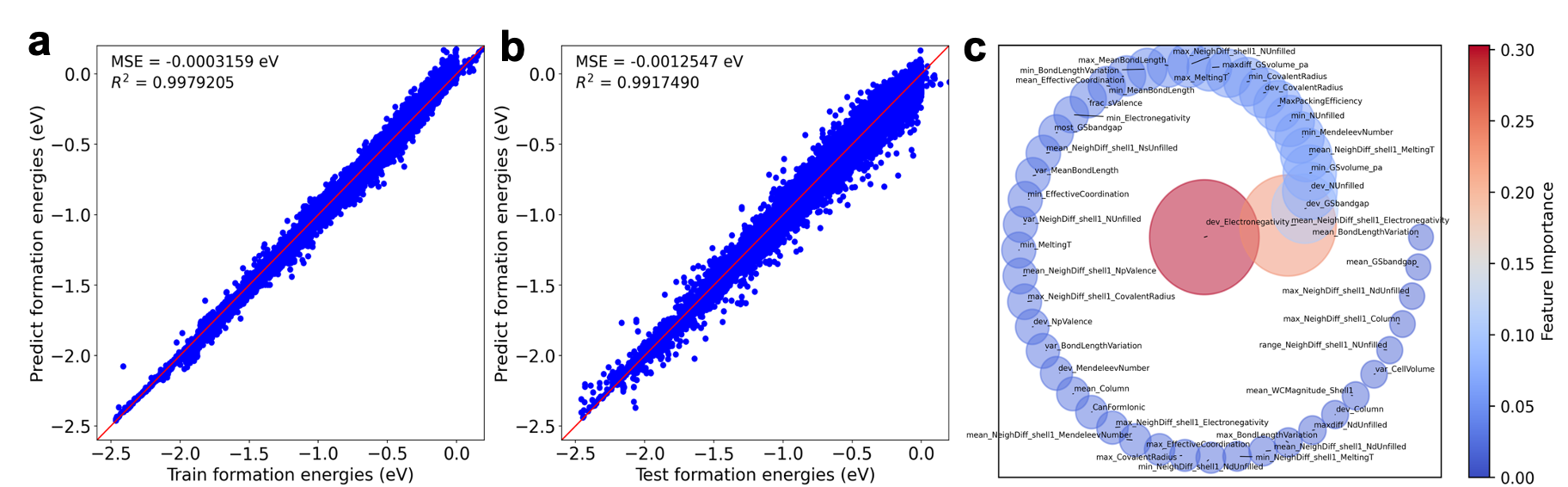}
    \caption{Performance evaluation of the trained stacked ensemble ML model using 126 structural features and 145 compositional features as descriptors for predicting the formation energy ($E_f$) of the mixed dataset: \textbf{a} 5,955 training set samples; \textbf{b} 2,551 test set samples. \textbf{c} Bubble plot representing the feature importance analysis results of the top 50 descriptors, where warmer bubble colors, larger bubble sizes, and positions closer to the center point indicate higher feature importance.}
    \label{ml}
\end{figure*}

We first extracted the structural information and the E$_f$ for all the S-based, Se-based, and Te-based candidates to construct a dataset for the ML training process.
As for the training process, we used Material-Agnostic Platform for Informatics and Exploration(magpie)~\cite{wolverton2016Magpie} to derive the structural features (126) and composition features (145) of chalcogenide as the descriptors, which were put into Autogluon together with the corresponding formation energy for training. 
Based on the three different bases, four different training sets were generated, including the S-based trainset, Se-based trainset, Te-based trainset, and Mixed trainset, which were used to train four different models for different scenarios. 
For the models of S-, Se-, and Te-based trainsets, the goal is to provide more accurate predictions for E$_f$ of Te-, Se-, and S-based chalcogenide structures, respectively. 
For the mixed trainset model, the goal is to provide predictions for regular chalcogenide structures. 
As shown in Fig.~\ref{ml}, we show the performance of the mixed model on the train and test sets, while the predictions of the remaining three models based on different elements are shown in Fig.~S5. 
We conducted a feature importance analysis, identifying the top 8 features with FI larger than 0.018.
These features include the mean absolute deviation between electronegativities of all atoms in the compound and the average electronegativity of the compound (mad$\_\chi$), the mean difference of electronegativity between neighbor atoms in the first nearest neighbor shell (md$\_\chi_{1st-shell}$), the mean absolute deviation of DFT ground state bandgap between all atoms and the average (mad$\_E_{g}$), the mean absolute deviation of the number of unfilled valence orbitals between all atoms and the average (mad$\_$UVO), minimum DFT-computed volume of ground state solid (min$\_V$), the mean difference of the melting temperature between neighbor atoms in the first nearest neighbor shell (md$\_T_{m}$), the minimum Mendeleev Number of the elements in the compound (min$\_M$), and the minimum of the number of unfilled valence orbitals (min$\_$UVO), as shown in the bubble plot of Fig.~\ref{ml}c.
It is noteworthy that the FI of electronegativity derived from composition information (0.092) and the structural information (0.071) contribute to more than 15\% of the ML model prediction.
Additionally, the number of unfilled valence orbitals is also highly significant.
The FI histogram of the top 50 features is shown in Fig.~S6, and a detailed explanation of each descriptor can be found in the supplementary of Ref.~\cite{ward2017structuralmagpie}.

\subsection{High-throughput Calculation of Band Gap}
According to the Shockley-Queisser (S-Q) efficiency limit, the band gap ($E_{g}$) of an ideal solar absorber in a single-junction cell should be close to 1.34 eV~\cite{shockley1961detailed}. 
Considering the efficiency in estimating the electronic $E_{g}$, a method with a low computational cost is required to accelerate the screening process for target candidates.
Therefore, after evaluating the thermodynamical stability of all ternary candidates, we then roughly calculated $E_{g}$ of 1417 stable phases and 8023 metastable phases using the approximate GGA-PBE exchange functional~\cite{perdew1996generalized}. 
Previously reported that GGA-PBE functional usually underestimates the $E_{g}$ by 0.5-0.6 eV~\cite{wang2019materials, li2019high}. 
Considering the subsequent investigation on diverse optoelectronic applications, a screening criterion of $E_{g} >$ 0 eV was set preliminary.
As a result, the number of ternary candidates is reduced to 6140.
Then, we performed high-throughput calculations based on the scPBE0 functional to obtain an accurate electronic band gap of the remaining 6140 semiconductors.
We made benchmarks on a series of experimentally-known materials, including CdTe, ZnX (X=O, S, Se, and Te), and many others.
The comparison between the experimental and theoretical $E_{g}$ calculated results is shown in Fig. S1b. 
As shown in Fig.~\ref{prototype}a, candidates with AgGaS$_{2}$ or LiGaS$_{2}$ structural prototype are more likely to exhibit direct bandgaps.
Both two prototypes consist of corner-sharing AX$_{4}$ and BX$_{4}$ tetrahedra, leading to a three-dimensional framework.
In addition to the appropriate electronic band gap, the gap difference ($\bigtriangleup E_{g}$) between the lowest direct band gap ($E_{_{g}}^{dir}$) and the fundamental band gap, which can be direct or indirect, acts as a key factor affecting the performance of the optoelectronic application. 
A smaller $\bigtriangleup E_{g}$ characteristic is beneficial to reduce the nonradiative recombination loss in the semiconductors, leading to higher efficiency for photovoltaic materials. 
Previously reported that semiconductors exhibiting indirect bandgap with $\bigtriangleup E_{g} <$ 0.2 eV can also be potentially promising solar cell materials\cite{yu2012identification}.
We selected chalcogenides with bandgaps in the range of 0.8-2.0 eV and $\bigtriangleup E_{g}$ less than 0.2 eV for further investigation of optoelectronic properties.
Accordingly, the number of candidates is reduced to 1477.
It should be noted that a suitable band gap is not sufficient to figure out promising photovoltaic materials, which are also detrimentally affected by intrinsic defects and impurities.
The induced localized trap states by defects can contribute to the nonradiative carrier recombination process, leading to efficiency loss.
However, with the consideration of the heavy computational cost of investigating the defect physics in semiconductors, related studies will be done in the future.

\subsection{Photovoltaic Performance and Electronic Structures}

\begin{figure}
    \centering
    \includegraphics[height=9 cm]{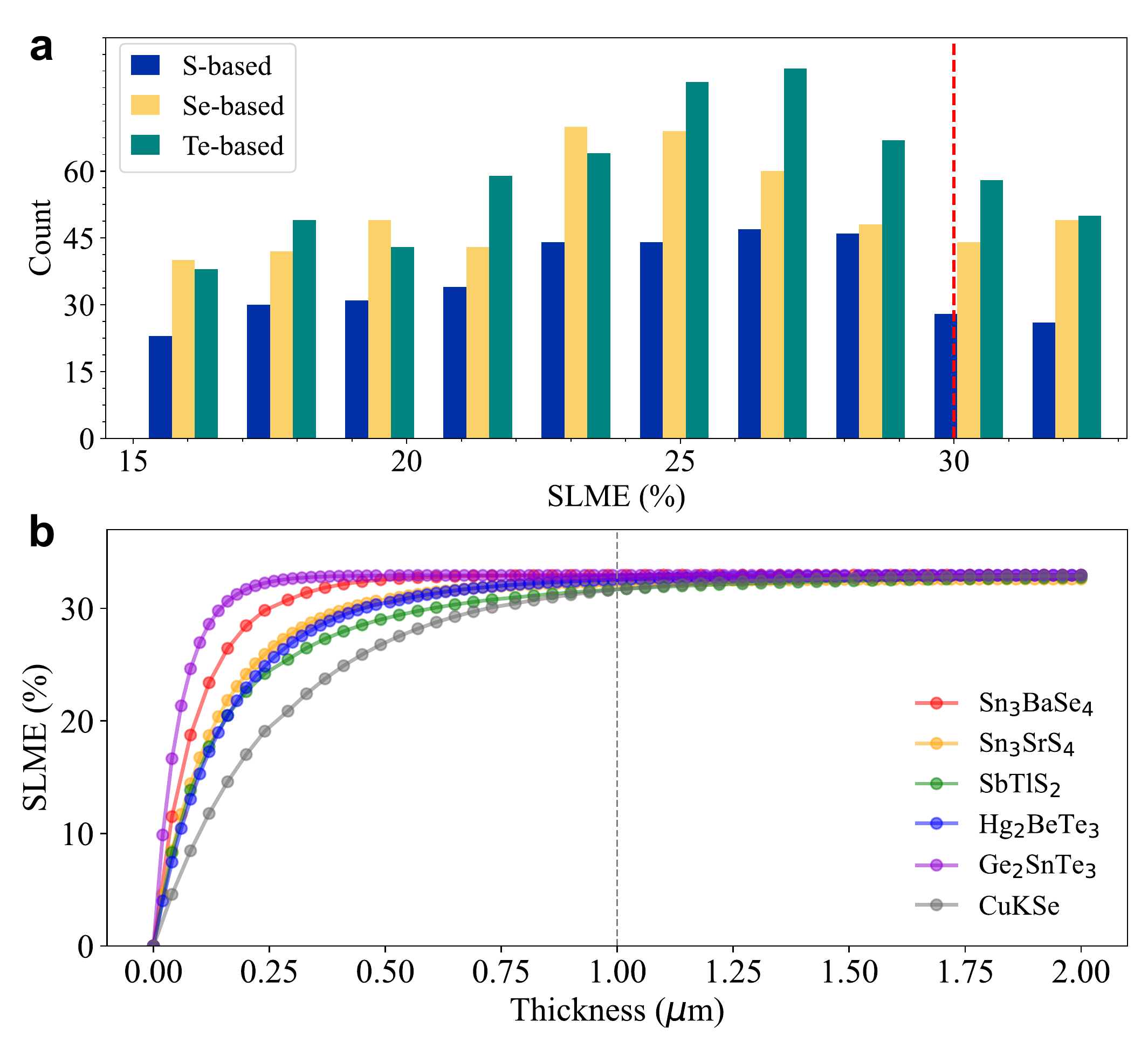}
    \caption{\textbf{a} Statistics on the number of candidates in different SLME intervals. S-, Se-, and Te-based candidates are colored with blue, yellow, and green, respectively. \textbf{b} Thickness-dependent SLME values of the top two compounds in S-, Se- and Te-based chalcogenides (SLME $\ge$ 32\%), respectively.}
    \label{optical}
\end{figure}

Since we have obtained the band gap of chalcogenides combined with satisfied stability, we can roughly evaluate their performances as photovoltaic materials based on the S-Q limit. 
For the solar cell material, there is a well-known selection criterion that a promising absorber generally relies on favorable ones with direct bandgap characteristics, which facilitates the strong optical absorption in solar cells with reduced thickness.
However, the conventional evaluation of S-Q efficiency is roughly dependent on the band gap value, failing to take into account the condition of the dipole-forbidden rather than a dipole-allowed direct transition in a few direct-gap materials.
Therefore, the selection criterion simply based on the S-Q efficiency has been increasingly proven to be insufficient to discover promising photovoltaic materials.
Accordingly, we adopted the "Spectroscopic Limited Maximum Efficiency (SLME)" as a target property to evaluate the efficiency of candidates\cite{yu2012identification}. 
We calculated the SLME for all candidates with bandgaps in a range of 0.8-2.0 eV. 
A typical film thickness of 2 $\mu$m in combination with the standard AM1.5G solar spectrum illumination was applied.
As shown in Fig. S8, the calculated SLME values of GaAs, CdTe, and CuGaSe$_{2}$ are in good agreement with experiments or references, indicating the reliability of our theoretical predictions.
There are 15, 32, and 35 candidates that exhibit theoretical efficiency of $\ge$ 32\% for S-, Se-, and Te-based chalcogenides, respectively (as shown in Fig.~\ref{optical}a). 
Such high efficiency of these candidates can be attributed to their suitable bandgaps with direct-allowed transitions, ultralow $\bigtriangleup E_{g}$, and high optical absorption. 

To investigate the photovoltaic performance, we further calculated the electronic structures and the carrier effective masses of candidates with SLME $>$ 32\%. 
For photovoltaic devices, high mobility of photogenerated charge carriers is required for the facilitated transport and efficient collection by electrodes.
A key factor determining carrier mobility is the carrier effective mass, which is calculated from the second derivatives of band structure curves near the valence band maximum (VBM, for holes) and conduction band minimum (CBM, for electrons), respectively.
After obtaining the band structures at the scPBE0 level and the carrier effective masses of candidates, we finally selected 65 promising materials for photovoltaic applications, as listed in Table S3. 
The band structures of 65 candidates are shown in Fig.~S8. 
From the results, all the candidates exhibit direct or quasi-direct band structures with reasonable carrier effective masses. 
A few candidates, such as BaTiSe$_{3}$ and Ag$_{2}$GeSe$_{3}$, exhibit weakly indirect band structures with extremely small $\bigtriangleup E_{g}$ less than 10 meV, which is lower than that of CH$_{3}$NH$_{3}$PbI$_{3}$ ($\sim$ 30 meV).
This quasi-direct band structure is beneficial to enhance the carrier lifetime, and simultaneously maintain high optical absorption.
Among 65 candidates, there are 52 compounds with ultra-low effective masses ($< m_{0}$) for both holes and electrons.
We selected the top two candidates with ultra-low carrier effective masses for S-, Se-, and Te-based chalcogenides, respectively, and calculated their thickness-dependent SLME values, as shown in Fig.~\ref{optical}b. 
The SLME values of these six candidates can almost saturate at ~1.0 $\mu$m, indicating their potential for thin-film solar cell absorbers.

\begin{figure}
    \centering
    \includegraphics[height=9 cm]{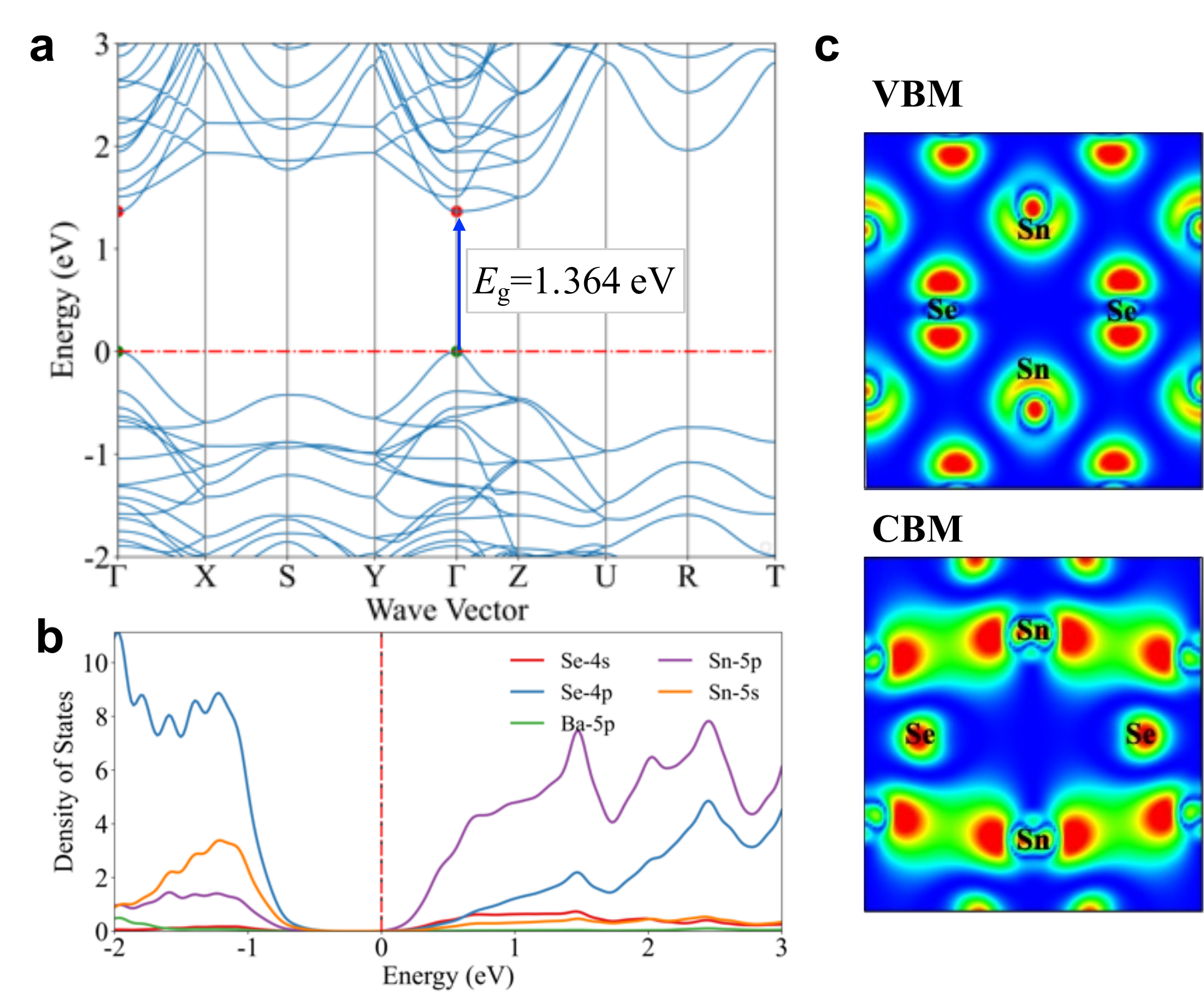}
    \caption{\textbf{a} Electronic band structure, \textbf{b} projected density of states, and (c) band composed charge densities of VBM (upper panel) and CBM (bottom panel) with isosurface level of 0.0002 for Sn$_{3}$BaSe$_{4}$.}
    \label{elecstru}
\end{figure}

To further investigate the physical mechanism of the candidates, we take Sn$_{3}$BaSe$_{4}$ as an example to study the orbital contribution near the band edges, as well as chemical bonding characteristics. 
As shown in Fig.~\ref{elecstru}, Sn$_{3}$BaSe$_{4}$ exhibits a direct band structure with $E_{g}$=1.364 eV, which is very close to the optimal band gap of the single-junction solar cell (1.34 eV).  
Both the VBM and CBM are located at $\Gamma$-symmetry point.
The calculated $m_{h}^{*}$ and $m_{e}^{*}$ are 0.08 $m_{0}$ and 0.20 $m_{0}$, respectively, indicating a strong mobility of carriers.
The top valence band dominantly consists of Sn-$s$ and Se-$p$ orbitals, while the bottom conduction band is composed of Sn-$p$ and Se-$p$ combined with Se-$s$ orbitals.
We also calculated the projected crystal orbital Hamilton populations, as shown in Fig.~S9.
The results show that the Sn-Se chemical bonding exhibits significant antibonding characteristics at both VBM and CBM, leading to a typical lone-pair Sn-5$s^{2}$-based electronic configuration in this compound.
Previous studies have demonstrated this n$s^{2}$-cation-containing electronic configuration as a physical origin introducing diverse superior optoelectronic properties in CH$_{3}$NH$_{3}$PbI$_{3}$, including bipolar carrier transport, strong light emission/absorption, defect tolerance, reduced carrier recombination, and many others~\cite{li2021alternative, yin2014unique, yin2017exploring, shi2019impact}. 
Similar conditions can also be found in other n$s^{2}$-cation-containing candidates, such as Ge$_{2}$SnTe$_{3}$ and SbTlS$_{2}$.
It suggests that our screened candidates represented by Sn$_{3}$BaSe$_{4}$ can potentially act as promising alternatives to Pb-based halide perovskites for optoelectronic applications, which requires further in-depth investigations.

\section{Discussion}
In summary, we have presented a comprehensive high-throughput screening process based on the idea of ion substitution and ab initio calculations to explore promising ternary chalcogenides for optoelectronic applications. 
By replacing A- and B-site cations in ternary A$_x$B$_y$X$_z$ chalcogenides crystallized in 34 structural prototypes, a rich material database containing  $>$ 400,000 A$_x$B$_y$X$_z$ chalcogenide candidates was constructed.
After a series of systematic first-principles calculations, we found 7943 unknown materials exhibiting satisfied phase stability against chemical decomposition.
On this basis, the chemical distribution of phase stability in terms of the structural prototype and chemical composition was established, respectively.
Furthermore, an effective stacked ensemble model based on machine learning algorithms was generated to predict the phase stability directly.
By using the developed self-consistent hybrid functional,
a photovoltaic-functionality-oriented material screening process involving $>$ 6000 stable and metastable candidates was performed, providing a final identification of 52 chalcogenides as promising absorbers in single-junction solar cells.
These materials exhibited good thermodynamic stabilities (E$_f < 0$ eV and $\Delta E_h <$ 50 meV/atom), suitable band gaps (1.1-1.5 eV), direct or quasi-direct electronic band structures ($\bigtriangleup E_{g} <$ 10 meV), ultra-low carrier effective masses ($<$ m$_{0}$), and high efficiencies ($>$ 32 \%).
In particular, a series of materials represented by Sn$_{3}$BaSe$_{4}$ show similar optical characteristics as CH$_{3}$NH$_{3}$PbI$_{3}$ owing to the intrinsic lone-pair n$s^{2}$-cation-containing electronic configuration, which has been verified as an origin for the high conversion efficiency of the solar cell devices fabricated by Pb-based halide perovskites. 
Our work provides a promising routine for exploring novel ternary chalcogenide semiconductors with superior performances for optoelectronic applications.
And the subsequent experimental efforts aiming at the synthesis and performance characterization are strongly called for.

\section{Methods}
The first-principles calculations were performed using an in-house developed HTP calculation environment~\cite{opahle2012high,opahle2013high}, as recently demonstrated for antiperovskites~\cite{singh2018high, singh2023high}, Heusler compounds~\cite{luo2022high, he2022computationally}, MAX phases~\cite{zhang2019high} and MAB phases~\cite{shen2021designing}. 
All the calculations are carried out within the DFT framework, as implemented in the Vienna ab initio simulation package~\cite{kresse1996efficient,kresse1999ultrasoft}. 
The interaction between core and valence electrons was explained using the frozen-core projector augmented wave pseudopotentials. 
In our calculations, the generalized gradient approximation with Perdew-Burke-Ernzerhof (GGA-PBE) functional was used as the exchange-correlation functional~\cite{perdew1996generalized}.
The crystal structures were fully optimized with the energy convergence criteria of 1 $\times$ 10$^{-5}$ eV.
The kinetic energy cutoff for plane-wave basis was set to 500 eV, in conjunction with a grid spacing of 2$\pi$ $\times$ 0.03 \AA$^{-1}$ for Brillouin zone sampling. 
To get a consistent and accurate convex hull, all the competing phases in the OQMD are recalculated using the same parameters in this work.
For the calculation of electronic structures, we considered the spin-orbital coupling (SOC) interaction for compounds containing heavy elements with an atomic number of more than 72. 
In order to obtain accurate electronic structures, we employed a {\it self-consistent} hybrid functional approach (hereafter denoted as ``scPBE0") to avoid evaluating the proportion of Hartree-Fock exact-exchange term ($\alpha$) empirically ~\cite{skone2014self}.
The carrier effective masses were calculated via BoltzTrap\cite{madsen2006boltztrap}.
The prediction of the high-accuracy formation energy of chalcogenides was performed using $ensembling$ multiple machine learning (ML) algorithms integrated into the AutoGluon-Tabular framework, which has demonstrated the efficiency in data processing, deep learning, and multi-layer model ensembling~\cite{erickson2020autogluon,qi2021autogluon,liu2022global}.
More computational methods on the scPBE0 functional  and photovoltaic conversion efficiency are provided in the Supplementary Information.

%\section*{Author Contributions}

\section*{Conflicts of interest}
The authors declare no competing financial interests.

\section*{Acknowledgements}
The Lichtenberg high-performance computer of the TU Darmstadt is gratefully acknowledged for the computational resources where the calculations were conducted for this project.

\bibliography{ref}

\end{document}

\title{Supplemental Material for \\
``Accelerated Screening of Ternary Chalcogenides for High-Performance Optoelectronic Materials"}

\author{Chen Shen\textit{$^{1}$}}
\author{Tianshu Li\textit{$^{1,*}$}}
\author{Teng Long\textit{$^{2}$}}
\author{Nuno Miguel Fortunato\textit{$^{1}$}}
\author{Fei Liang\textit{$^{2}$}}
\author{Mian Dai\textit{$^{1}$}}
\author{Jiahong Shen\textit{$^{3}$}}
\author{Chris Wolverton\textit{$^{3}$}}
\author{Hongbin Zhang\textit{$^{1}$}}

\affiliation{\textit{$^{1}$Institute of Materials Science, Technical University of Darmstadt, Otto-Berndt-Str. 3, Darmstadt, 64287, Hessen, Germany.}}
\affiliation{\textit{$^{2}$School of Materials Science and Engineering, Shandong University, Jingshi Road 17923, Jinan, 250061, Shandong, China.}}
\affiliation{\textit{$^{3}$Department of Materials Science and Engineering, Northwestern University, Clark Str. 633, Illinois, 60208, Evanston, USA.}}

\maketitle

\section{Computational Methods of the Self-Consistent Hybrid Functional}
In order to avoid evaluating the proportion of Hartree-Fock exact-exchange term ($\alpha$) empirically~\cite{skone2014self},
the accurate electronic structures of thermodynamically stable chalcogenides were calculated using the so-called {\it self-consistent} hybrid functional approach (hereafter denoted as ``scPBE0").
For the typical hybrid functional, a portion of the local exchange-correlation potential is replaced by Hartree-Fock exact-exchange terms as follows,
\begin{equation} \label{eq1}
    v_{xc}(\mathbf{r},\mathbf{r}')=\alpha v_{_{x}}^{ex}(\mathbf{r},\mathbf{r}')+(1-\alpha)v_{x}(\mathbf{r})+v_{c}(\mathbf{r})
\end{equation}
In our work, the fraction $\alpha$ admixed in this equation was determined by the inverse of the static dielectric constant according to the screening behavior of nonmetallic system~\cite{alkauskas2011defect, marques2011density},
\begin{equation} \label{eq2}
    \alpha =\frac{1}{\epsilon_{\infty} } 
\end{equation}
which can be computed in a self-consistent cycle, as illustrated in Fig.S1a~\cite{skone2014self,gerosa2015electronic}.

\begin{figure*}
    \centering
    \includegraphics[height=7 cm]{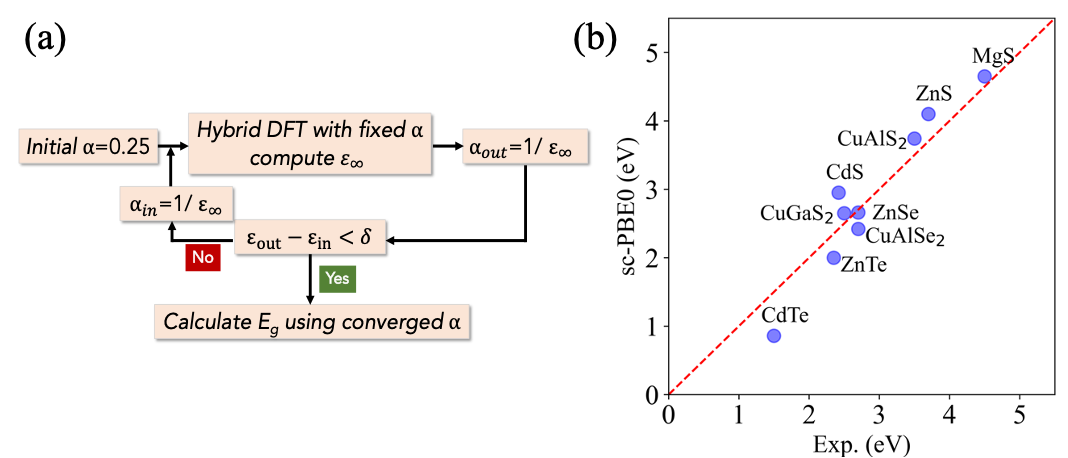}
    \caption{(a) Diagram of the self-consistent hybrid functional scheme. (b) The comparison of electronic band gaps between the results calculated via sc-PBE0 functional and experimental band gaps.}
    \label{figs1}
\end{figure*}

\section{Computational Methods of the "Spectroscopic Limited Maximum Efficiency"}
The standard AM 1.5G solar spectrum at 25 $^{\circ}$C is used for simulating the efficiency. The maximum efficiency of each candidate with a band gap in the range of 0.8 - 2.0 eV is calculated on the basis of the optical absorption coefficient with modified Shockley-Queisser methods.
The short-circuit voltage ($J_{SC}$) is calculated by,
\begin{equation}
\label{Jsc}
    J_{SC} = e\int^{\infty}_{0}\alpha(E)I_{sun}(E)dE
\end{equation}
where $\alpha(E)$ is optical absorption coefficient at photonic energy E, $I_{sun}(E)$ is the sunlight photon flux at energy E, and e is the fundamental electron charge.
The $J-V$ curve of an ideal $pn$ junction follows the formula,
\begin{equation}
\label{JV}
    J = J_{SC} - J_{0}(1 - exp(eV/kT))
\end{equation}
Without the consideration of nonradiative electron-hole (e-h) recombination, the radiative  e-h recombination dominantly contributes to $J_{0}$ part. 
Once we obtain the optical absorption coefficient $\alpha(E)$ for a material, its maximum quantum efficiency can be calculated by numerically maximizing J$\cdot$V according to Eq.~\ref{JV} and dividing the maximized J$\cdot$V by the total AM1.5G solar power.

\begin{table*}
\caption{Space group, space group number, isostructural prototype compounds, the numbers of S-/Se-/Te-based candidates with $\Delta E_h <$ 50 meV ($N_{\Delta E_h<50}$), band gap $E_{g} >$ 0 eV ($N_{E_g>0}$), and direct band structure ($N_{d\_Eg}$) in 34 structural prototypes, respectively. All the prototypes are available in open material databases.}
\label{tabs1}
\begin{tabular*}{\textwidth{}}{@{\extracolsep{\fill}}ccccccc}
\hline
\textbf{Index} & \textbf{Space Group} & \textbf{\makecell[c]{Space Group \\ Number}} & \textbf{\makecell[c]{Isostructural \\ Prototype}} & \textbf{\makecell[c]{$N_{\Delta E_h<50}$ \\ (S/Se/Te)}} & \textbf{\makecell[c]{$N_{E_g>0}$ \\ (S/Se/Te)}} & \textbf{\makecell[c]{$N_{d\_Eg}$ \\ (S/Se/Te)}} \\
\hline
1  & Cmc2_1   & 36  & Ag$_2$GeS$_3$  & 112/111/93  & 64/55/45    & 34/26/24   \\
2  & Pmn2_1   & 31  & Ag$_3$PS$_4$   & 122/135/82  & 56/51/34    & 27/17/19   \\
3  & I$\Bar{4}$2d    & 122 & AgGaS_2   & 108/105/95  & 88/79/77    & 75/63/65   \\
4  & Pnma    & 62  & BaZrS_3   & 64/80/82    & 56/60/52    & 16/21/7    \\
5  & P6_3/mmc & 194 & BaZrSe_3  & 10/21/12    & 3/0/0       & 1/0/0      \\
6  & P2_1/c   & 14  & BiOsSe   & 254/313/304 & 180/191/167 & 45/52/53   \\
7  & P2_1/c   & 14  & Ca$_2$Sb$_2$S$_5$ & 97/105/85   & 83/86/66    & 24/27/19   \\
8  & I$\Bar{4}$2d    & 122 & CdCe$_2$S$_4$  & 115/128/70  & 61/40/16    & 24/7/1   \\
9  & I4/mcm  & 140 & CeTlTe_2  & 17/30/42    & 15/23/26    & 8/11/10 \\
10 & Fd$\Bar{3}$m    & 227 & CoCr$_2$S$_4$  & 147/149/94  & 47/33/6     & 14/14/3    \\
11 & P3m1    & 156 & CsBiSe_2  & 72/63/82    & 32/34/40    & 5/7/7   \\
12 & P2_1/c   & 14  & CsGaSe_3  & 156/211/194 & 145/176/145 & 29/37/12   \\
13 & R$\Bar{3}$m     & 166 & CsPrS$_2$   & 197/223/252 & 115/111/95  & 12/21/19   \\
14 & P2_1/c   & 14  & CsSbSe_2  & 263/235/252 & 196/156/147 & 64/54/54   \\
15 & Cc      & 9   & Cu$_2$GeS$_3$  & 145/177/165 & 92/98/81    & 44/48/35   \\
16 & I$\Bar{4}$2m    & 121 & Cu$_3$SbS$_4$  & 83/88/63    & 33/30/20    & 20/20/18   \\
17 & Pnma    & 62  & CuAsS    & 134/154/155 & 104/102/81  & 44/51/38   \\
18 & F$\Bar{4}$3m    & 216 & HfTeRu   & 16/20/18    & 12/14/12    & 2006/8/8   \\
19 & Pnma    & 62  & K$_3$NbSe$_4$  & 122/143/87  & 74/87/60    & 42/41/24   \\
20 & P4/nmm  & 129 & KAgSe    & 58/73/100   & 20/24/26    & 6/11/10 \\
21 & P4/mmm  & 123 & KBiSe$_2$   & 46/58/68    & 28/26/22    & 7/12/15 \\
22 & F$\Bar{4}$3m    & 216 & KNaSe    & 5/6/32      & 0/0/8       & 0/0/6      \\
23 & P2_1/c   & 14  & LaSbSe   & 154/180/219 & 81/81/77    & 36/36/37   \\
24 & Cmcm    & 63  & LaYbSe$_3$  & 65/55/64    & 36/26/28    & 14/9/3   \\
25 & Pna2_1   & 33  & LiGaS$_2$   & 158/195/165 & 129/145/128 & 64/67/92   \\
26 & P$\Bar{3}$m1    & 164 & NaTiS$_2$   & 163/202/227 & 83/104/97   & 1/0/2      \\
27 & C2/m    & 12  & Nb$_2$CrSe$_4$ & 244/355/321 & 79/88/40    & 8/4/0      \\
28 & Pnma    & 62  & Rb$_2$MoSe$_4$ & 57/63/32    & 35/32/11    & 17/19/6    \\
29 & Cmcm    & 63  & RbAuSe   & 48/65/72    & 36/42/39    & 13/23/27   \\
30 & R3m     & 160 & RbTiS$_2$   & 99/109/106  & 47/47/41    & 3/6/4   \\
31 & P2_13    & 198 & SbIrSe   & 85/112/146  & 55/50/61    & 9/13/18    \\
32 & I4_{1}/amd & 141 & ScUS$_2$    & 74/98/92    & 38/55/45    & 23/32/25   \\
33 & I4/mmm  & 139 & UGeTe    & 31/29/42    & 8/7/3    & 2/2/1   \\
34 & P4/nmm  & 129 & ZrSiS    & 52/66/102   & 17/18/25    & 8/9/8  \\
\hline
\end{tabular*}
\end{table*}

%\section{Properties of 34 structural prototypes}
\begin{table*}
  \caption{Statistics of the known and predicted ternary chalcogenides, categorized by the thermodynamic stability of  $A_xB_yCh_z$ phases. All spaces are categorized by the $\Delta E_f$ of the ternary chalcogenides with the lowest formation energy. Metastable phases are categorized by their energy above the convex hull, $\Delta E_h$}
  \label{tstability}
\begin{tabular*}{\textwidth}{@{\extracolsep{\fill}}llll}
\hline
\textbf{Stable $A_xB_yCh_z$ phases}     & \textbf{Previously know} & \textbf{Newly predicted} & \textbf{Total number} \\
S-based                           &        112                  &             485             &               597        \\
Se-based                          &        80                 &             549            &                629       \\
Te-based                          &        33                 &             548             &               581        \\
In total                          &        225                &             1192             &              1417         \\
\textbf{\makecell[l]{Metastable $A_xB_yCh_z$ phases \\($\Delta E_h < $ 50 meV)}} & \textbf{Previously know} & \textbf{Newly predicted} & \textbf{Total number} \\
S-based                           &           20               &               2455           &          2475             \\
Se-based                          &           47               &               2815           &          2862             \\
Te-based                          &           13               &               2673           &          2686             \\
In total                          &           80               &               7943          &       8023  \\
\hline
\end{tabular*}
\end{table*}

\begin{figure*}
    \centering
    \includegraphics[height=13 cm]{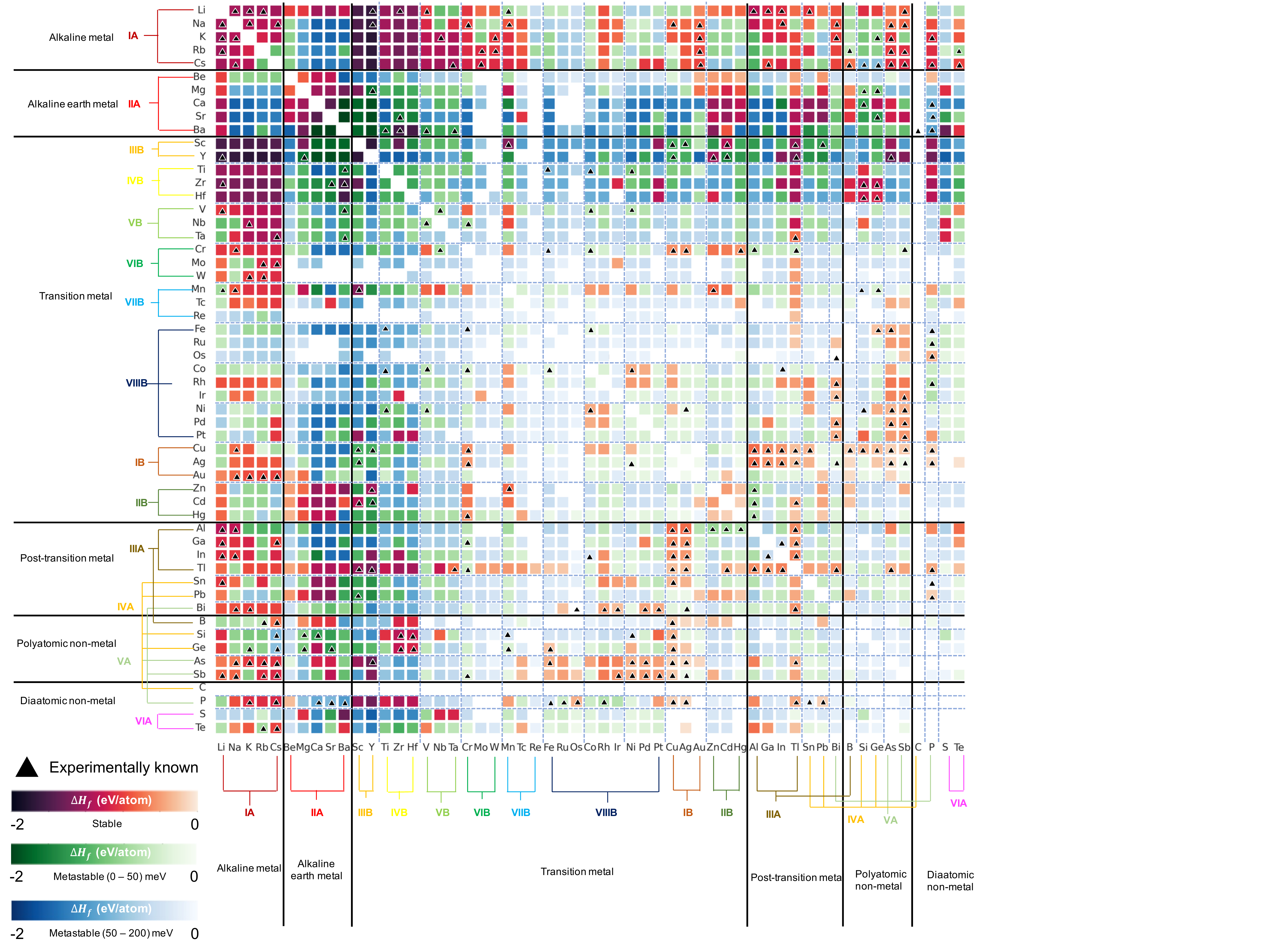}
    \caption{A map of the Se-based ternary chalcogenides, colored to represent the thermodynamic stability of the candidates with the lowest formation energy.}
    \label{figs3}
\end{figure*}

\begin{figure*}[c]
    \centering
    \includegraphics[height=13 cm]{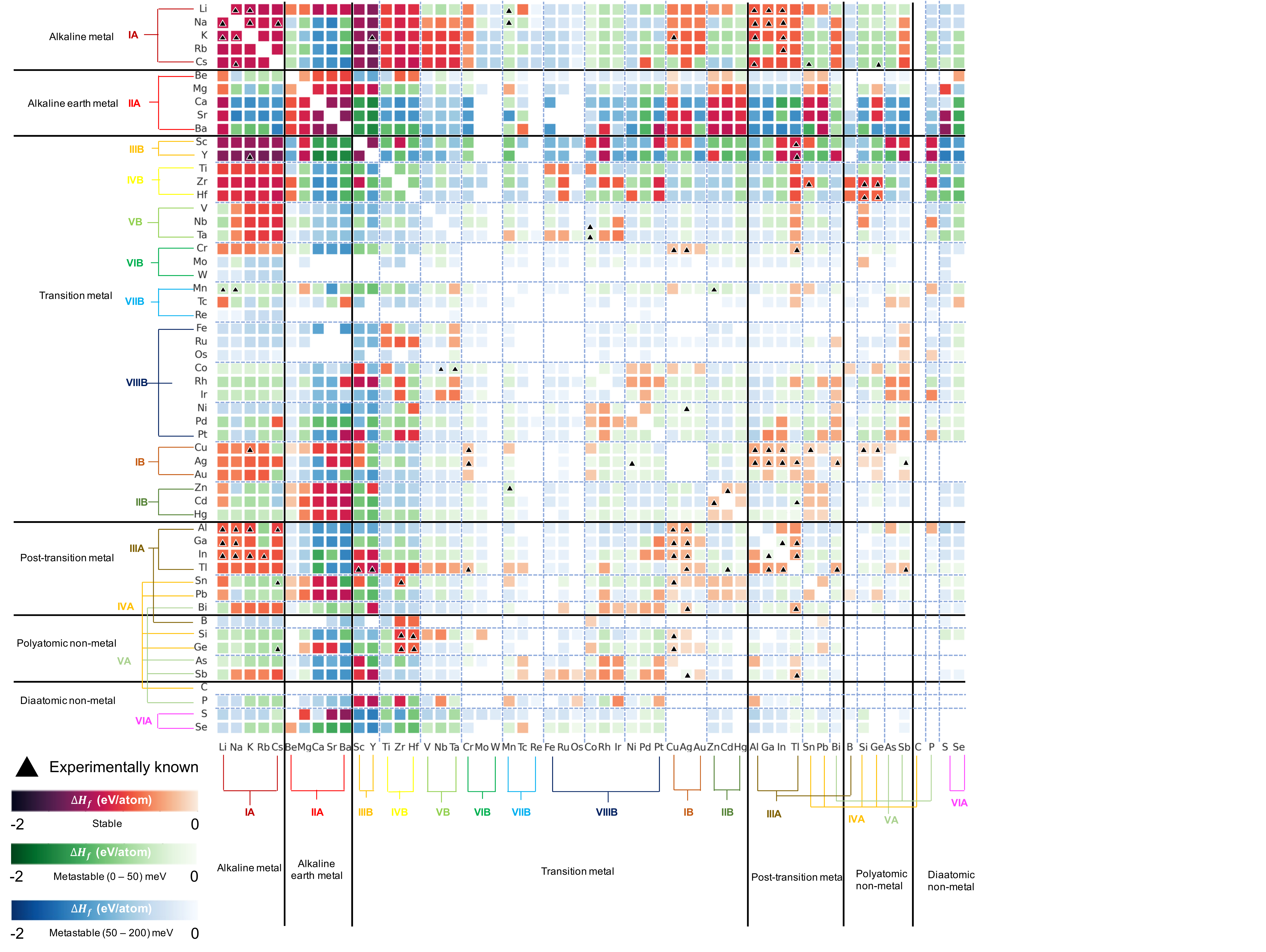}
    \caption{A map of the Te-based ternary chalcogenides, colored to represent the thermodynamic stability of the candidates with the lowest formation energy.}
    \label{figs4}
\end{figure*}

\begin{figure*}[c]
    \centering
    \includegraphics[height=13 cm]{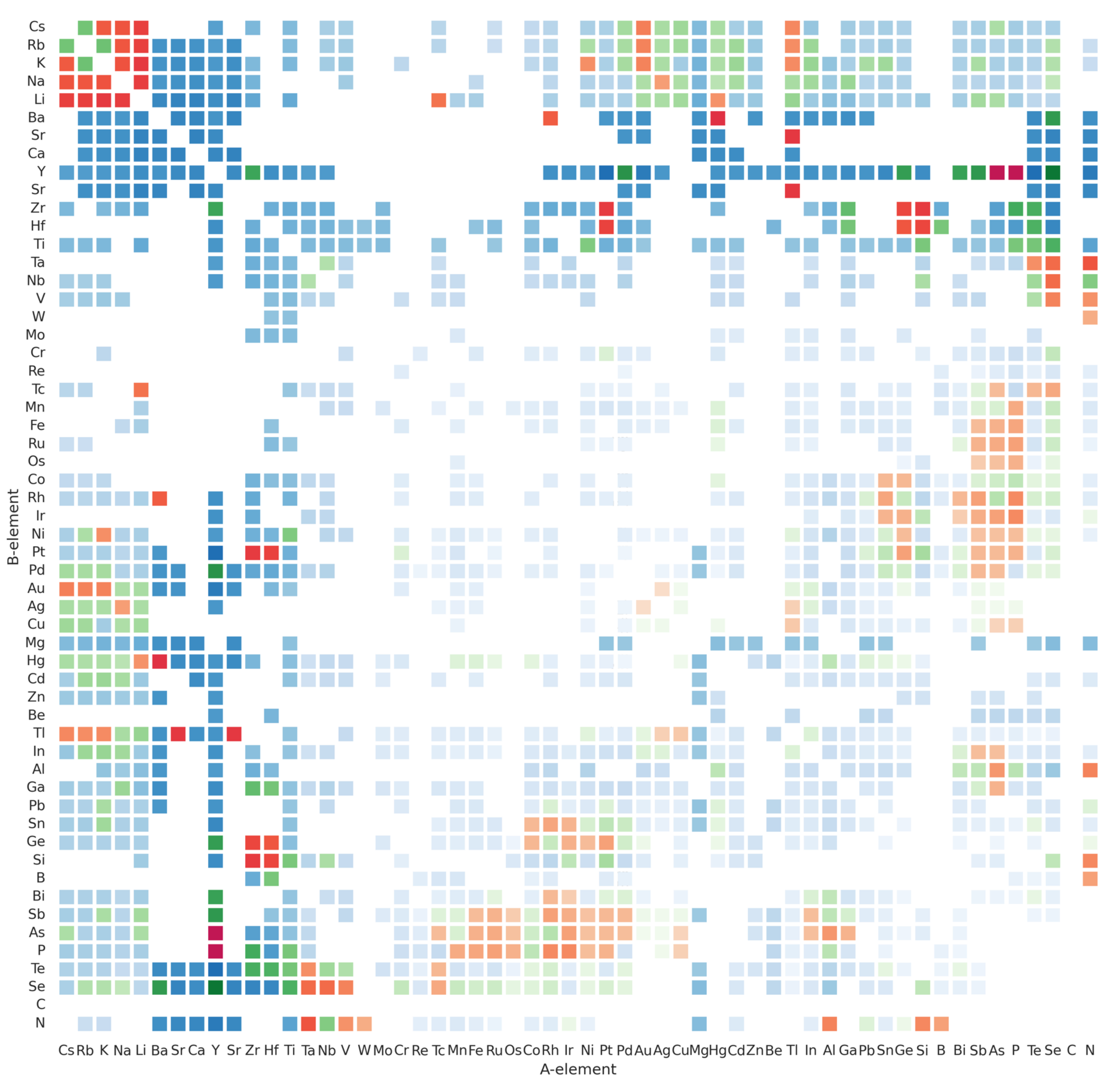}
    \caption{A map of the S-based ternary chalcogenides in a chemical formula of $A_{1}B_{1}Ch_{1}$, colored to represent the thermodynamic stability of the candidates with the lowest formation energy.}
    \label{figs4}
\end{figure*}

\begin{figure*}
    \centering
    \includegraphics[height=22 cm]{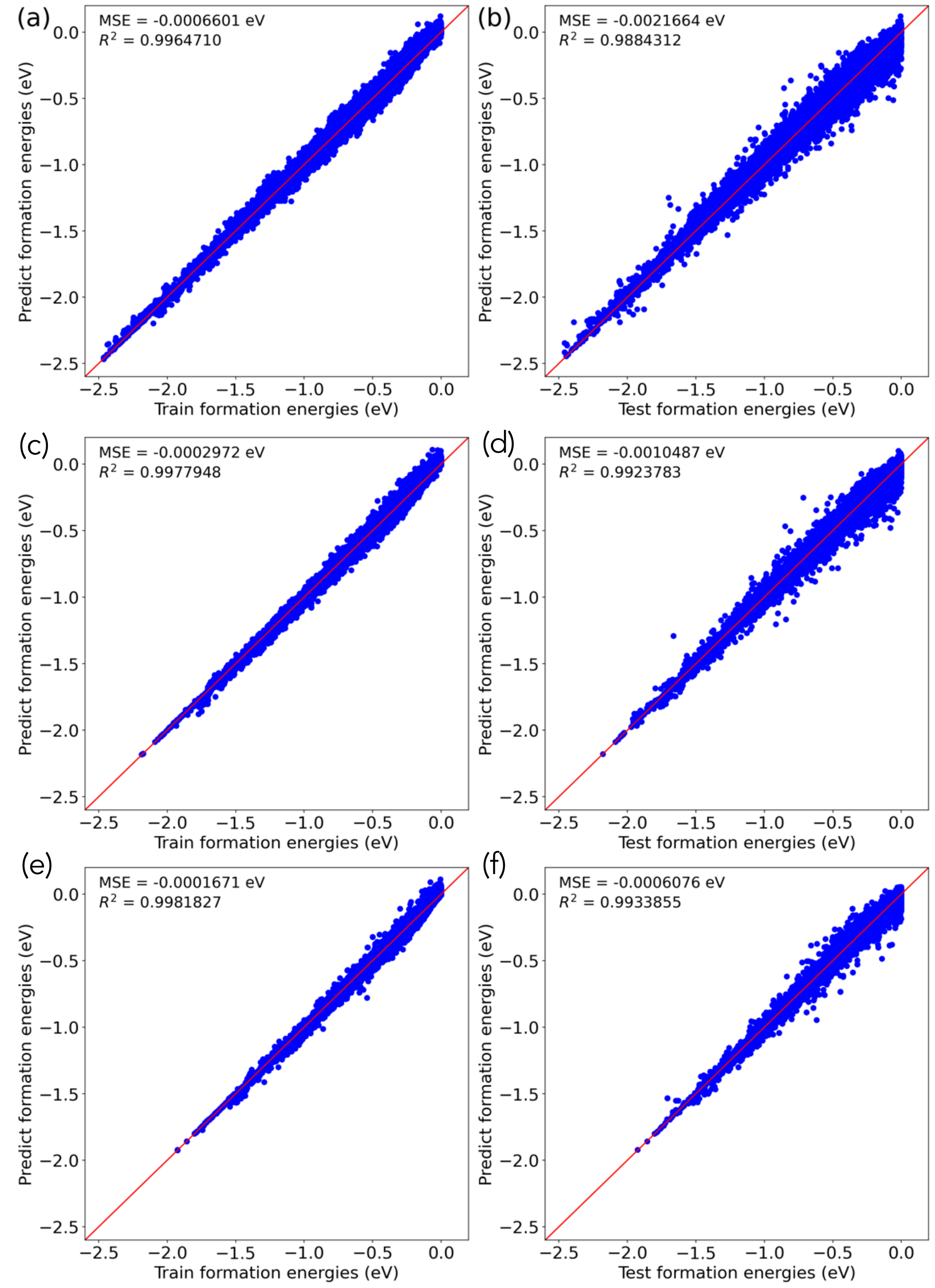}
    \caption{Performance evaluation of trained stack ensembling ML model on (a,c,e) train sets; (b,d,f) test sets of S-, Se- and Te-based chalcogenides, respectively.}
    \label{ml}
\end{figure*}

\begin{figure*}
    \centering
    \includegraphics[height=12 cm]{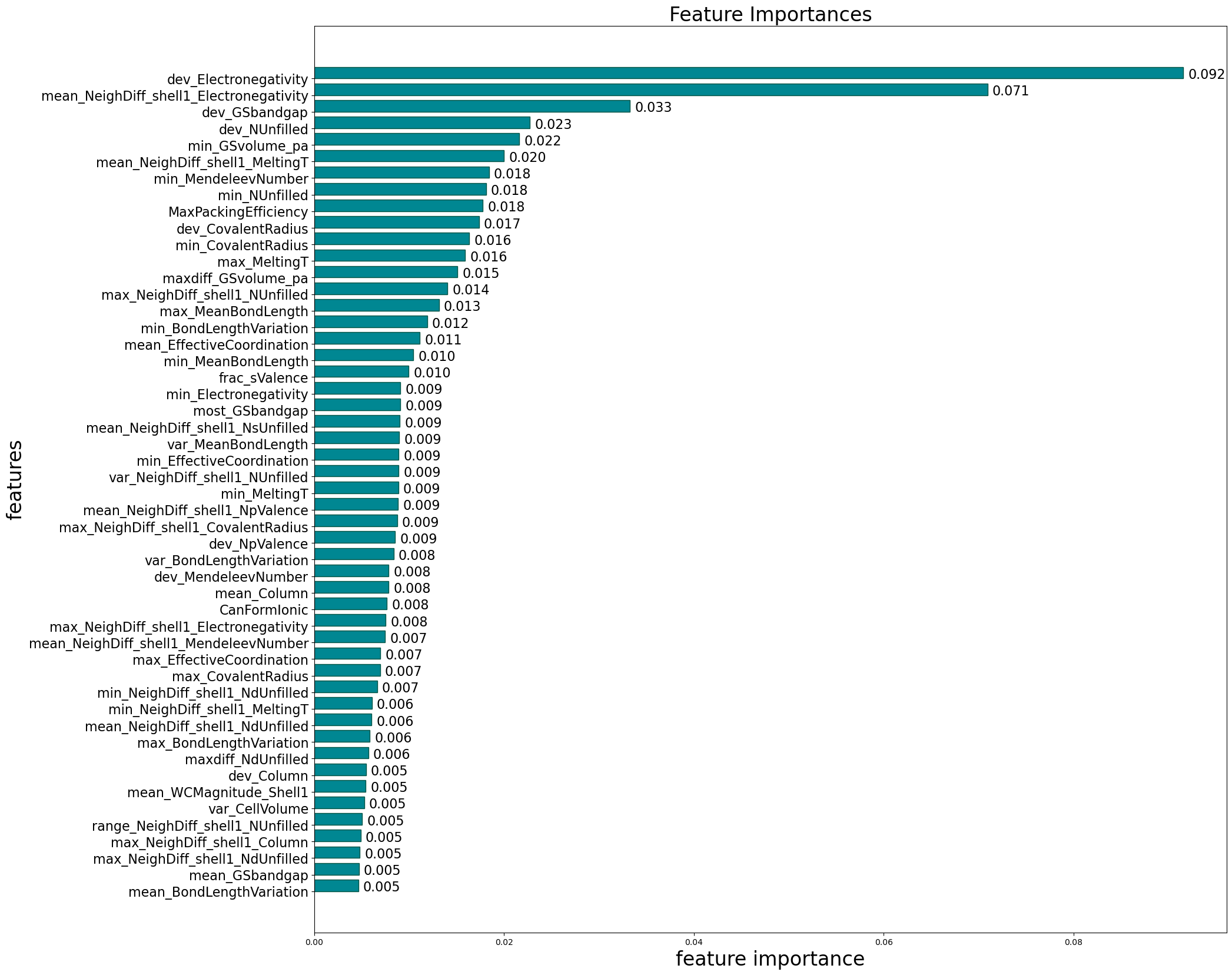}
    \caption{Feature importance analysis.}
    \label{FI}
\end{figure*}

\begin{figure*}
    \centering
    \includegraphics[height=7.5 cm]{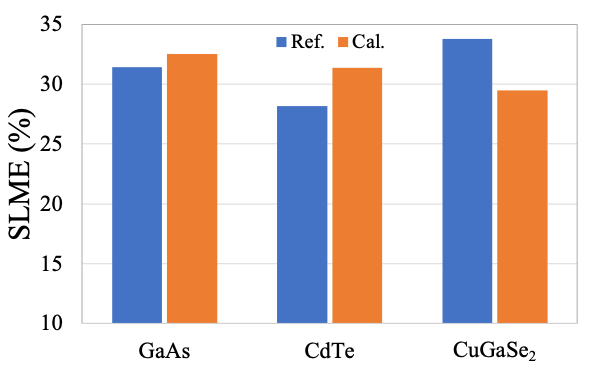}
    \caption{Referenced (Ref.) and calculated (Cal.) conversion efficiency of GaAs, CdTe and CuGaSe_{2}.}
    \label{figs2}
\end{figure*}

\begin{table*}
\centering
\caption{Structural prototype index according to Table~\ref{tabs1}, formation energy $E_f$ (eV/atom), the energy above the convex hull $\Delta E_h$ (eV/atom), electronic band gap at self-consistent hybrid functional level $E^{scPBE0}_{g}$ (eV), the difference between direct and indirect bandgaps $\Delta E_{g}$ (meV), electron effective mass ($m_{e}^{*}$), hole effective mass ($m_{h}^{*}$), and room temperature (298 K) SLME (\%) for thin-film solar cells with a thickness of 2 \mu m}
\label{tabs2}
\begin{tabular*}{\textwidth{}}{@{\extracolsep{\fill}}ccccccccc}
\hline
\textbf{$A_{x}B_{y}Ch_{z}$}    & \makecell[c]{Prototype \\ index} & \makecell[c]{$E_{f}$ \\ (eV/atom)} & \makecell[c]{$\Delta E_{h}$ \\ (eV/atom)} & \makecell[c]{$E^{scPBE0}_{g}$ \\ (eV)} &  \makecell[c]{$\Delta E_{g}$ \\ (meV)}    & $m_{e}^{*}$ & $m_{h}^{*}$ & SLME (\%) \\
\hline
Li$_2$Te$_2$Se$_5$ & 7               & -0.382       & 0.048     & 1.3539 & 0      & 1.21 & 0.12 & 32.96    \\
Sn$_3$BaSe$_4$  & 19              & -0.821       & 0.017     & 1.364  & 0      & 0.08 & 0.2  & 32.96    \\
CuKSe     & 6               & -0.637       & 0.027     & 1.3434 & 0      & 0.1  & 0.1  & 32.94    \\
Ge$_2$SnTe$_3$  & 15              & -0.141       & 0.015     & 1.3836 & 0      & 0.6  & 0.18 & 32.94    \\
Hg$_2$BeTe$_3$  & 15              & -0.196       & 0.029     & 1.3658 & 0      & 0.39 & 0.04 & 32.94    \\
GeSrTe$_3$   & 12              & -0.648       & -0.014    & 1.3614 & 0      & 0.21 & 0.17 & 32.94    \\
BaTiSe$_3$   & 12              & -1.502       & 0.012     & 1.365  & 0.546 & 0.6  & 0.5  & 32.93    \\
FeSeS     & 31              & -0.394       & 0.035     & 1.3257 & 0      & 1.55 & 0.38 & 32.83    \\
SbTlS$_2$    & 30              & -0.311       & 0.026     & 1.3381 & 0      & 0.35 & 0.17 & 32.73    \\
CsTlTe$_2$   & 9               & -0.54        & 0.043     & 1.398  & 0      & 0.89 & 0.31 & 32.72    \\
Ag$_2$GeSe$_3$  & 15              & -0.195       & 0         & 1.4049 & 0.562 & 0.26 & 0.14 & 32.69    \\
WLi$_2$Se$_4$   & 10              & -0.591       & 0         & 1.3579 & 0      & 0.74 & 0.37 & 32.66    \\
CdPbTe$_3$   & 4               & -0.207       & 0.01      & 1.3648 & 9.008 & 1.05 & 0.19 & 32.66    \\
RhYTe     & 31              & -1.108       & 0         & 1.4109 & 0      & 0.12 & 0.13 & 32.63    \\
Ge$_3$PbTe$_4$  & 19              & -0.131       & 0.039     & 1.1811 & 0      & 0.56 & 0.17 & 32.61    \\
PbZnTe$_3$   & 4               & -0.195       & 0.042     & 1.1668 & 0      & 0.16 & 0.14 & 32.61    \\
Ag$_3$SbS$_4$   & 16              & -0.111       & 0.012     & 1.1658 & 0      & 0.65 & 0.06 & 32.6     \\
Sn$_3$SrS$_4$   & 19              & -0.851       & 0.021     & 1.3127 & 0      & 0.08 & 0.23 & 32.6     \\
Ge$_3$CaS$_4$   & 19              & -0.692       & 0.04      & 1.1566 & 0      & 0.28 & 0.1  & 32.6     \\
Ag$_2$SnTe$_3$  & 1               & -0.131       & 0.01      & 1.1472 & 0      & 0.63 & 0.09 & 32.6     \\
SrHfSe$_3$   & 4               & -1.591       & 0.044     & 1.3724 & 0      & 0.15 & 0.3  & 32.57    \\
Hf$_2$HgSe$_4$  & 27              & -1.139       & 0.017     & 1.4049 & 1.826 & 0.2  & 0.22 & 32.57    \\
ZnIr$_2$Se$_4$  & 8               & -0.368       & 0.047     & 1.2041 & 0      & 0.6  & 0.25 & 32.52    \\
KVSe$_3$     & 12              & -0.808       & 0.046     & 1.3699 & 0      & 1.13 & 2.14 & 32.52    \\
ZnIr$_2$Se$_4$  & 10              & -0.368       & 0.047     & 1.2065 & 0.241 & 0.8  & 0.32 & 32.5     \\
BiInTe    & 23              & -0.227       & 0.002     & 1.1323 & 0      & 0.74 & 0.3  & 32.49    \\
CdIr$_2$Se$_4$  & 8               & -0.36        & 0.03      & 1.2246 & 0      & 0.56 & 0.27 & 32.42    \\
GeSc$_2$Se$_4$  & 10              & -1.238       & 0.054     & 1.3613 & 17.152 & 0.65 & 0.54 & 32.38    \\
Cd$_2$SrTe$_4$  & 28              & -0.492       & 0.026     & 1.2318 & 0      & 0.17 & 0.18 & 32.38    \\
ZnIr$_2$Te$_4$  & 8               & -0.271       & 0.049     & 1.3407 & 18.770  & 0.87 & 0.28 & 32.35    \\
Li$_2$Zn$_2$Te$_5$ & 7               & -0.425       & 0         & 1.2415 & 0      & 0.48 & 0.04 & 32.34    \\
Zr$_2$TeSe$_3$  & 15              & -1.236       & 0.003     & 1.3463 & 19.387 & 0.15 & 0.21 & 32.32    \\
Pb$_3$BaTe$_4$  & 19              & -0.725       & 0.029     & 1.4361 & 0      & 0.72 & 0.12 & 32.32    \\
Hg$_2$MgTe$_3$  & 1               & -0.381       & 0.018     & 1.3121 & 0      & 0.87 & 0.04 & 32.32    \\
Ag$_2$SiSe$_3$  & 15              & -0.257       & 0.022     & 1.1945 & 0      & 0.28 & 0.1  & 32.3     \\
NbKSe$_3$    & 12              & -0.902       & 0.045     & 1.3902 & 0      & 1.14 & 1.27 & 32.27    \\
Cs$_3$TlS$_4$   & 19              & -0.757       & 0.042     & 1.3666 & 4.783 & 0.9  & 1.64 & 32.27    \\
AgGaTe$_2$   & 14              & -0.237       & 0.05      & 1.1641 & 0      & 0.74 & 0.07 & 32.27    \\
Sn$_3$BaS$_4$   & 19              & -0.843       & 0.028     & 1.4416 & 0      & 0.39 & 0.22 & 32.25    \\
GeSrTe$_2$   & 32              & -0.813       & -0.02     & 1.3087 & 0      & 0.08 & 0.12 & 32.25    \\
TiBaSe$_3$   & 12              & -1.502       & 0.013     & 1.4429 & 0      & 0.5  & 0.79 & 32.23    \\
Hg$_2$MgTe$_3$  & 15              & -0.386       & 0.013     & 1.2445 & 0      & 0.87 & 0.04 & 32.22    \\
Hg$_2$BeTe$_3$  & 1               & -0.2         & 0.025     & 1.4091 & 0      & 0.91 & 0.04 & 32.21    \\
Zr$_2$HgSe$_4$  & 27              & -1.144       & 0.017     & 1.193  & 3.460 & 0.26 & 0.26 & 32.19    \\
Li$_2$PbSe$_3$  & 15              & -0.689       & 0.03      & 1.265  & 0.126 & 0.47 & 0.2  & 32.18    \\
AuGaSe$_2$   & 25              & -0.338       & 0.053     & 1.3026 & 0      & 2.53 & 0.12 & 32.18    \\
CdBi$_2$Te$_4$  & 8               & -0.252       & 0.052     & 1.1625 & 10.695 & 0.37 & 0.07 & 32.18    \\
Sn$_3$CaTe$_4$  & 19              & -0.618       & 0.011     & 1.1042 & 0      & 0.1  & 0.16 & 32.16    \\
Hg$_3$BeS$_4$   & 16              & -0.319       & 0.036     & 1.4089 & 0      & 0.65 & 0.1  & 32.14    \\
MoLi$_2$Se$_4$  & 10              & -0.594       & 0         & 1.4498 & 0      & 0.3  & 0.27 & 32.13    \\
AgInSe$_2$   & 25              & -0.42        & 0.004     & 1.2979 & 0      & 0.32 & 0.05 & 32.12    \\
Ca$_2$GeTe$_4$  & 28              & -0.762       & 0.051     & 1.2664 & 0      &  1.08    &  1.70    & 32.12    \\
CdIr$_2$Te$_4$  & 8               & -0.277       & 0.041     & 1.3321 & 25.843 & 3.01 & 0.36 & 32.1     \\
SnSrTe$_2$   & 21              & -0.994       & 0         & 1.103  & 0      & 0.21 & 0.24 & 32.1     \\
LiAgSe    & 23              & -0.632       & 0.047     & 1.159  & 0      & 0.12 & 0.12 & 32.07    \\
TiRuTe    & 22              & -0.559       & 0         & 1.4162 & 0      & 0.3  & 0.63 & 32.07    \\
BaHfSe$_3$   & 4               & -1.622       & 0.007     & 1.2621 & 0      & 0.52 & 0.2  & 32.06    \\
RbTlTe$_2$   & 9               & -0.521       & 0.039     & 1.1072 & 0      & 1.21 & 0.25 & 32.06    \\
HgZrSe$_2$   & 13              & -1.005       & 0.011     & 1.4531 & 0      & 0.22 & 0.24 & 32.04    \\
Cd$_2$HgTe$_3$  & 1               & -0.352       & 0.009     & 1.2387 & 0      & 0.55 & 0.05 & 32.04    \\
Ag$_2$SnSe$_3$  & 1               & -0.237       & 0.006     & 1.2872 & 0      & 1.46 & 0.1  & 32.03    \\
Pb$_3$SrTe$_4$  & 19              & -0.736       & 0.006     & 1.256  & 0      & 0.21 & 0.06 & 32.03    \\
AgYTe$_2$    & 14              & -0.895       & 0.038     & 1.4166 & 0      & 0.58 & 0.2  & 32.01    \\
KTlTe$_3$    & 24              & -0.419       & 0.045     & 1.4614 & 0      & 1.5  & 0.18 & 32       \\
CaMgTe$_3$   & 12              & -0.883       & 0.021     & 1.4607 & 0      & 1.19 & 0.1  & 32       \\
\hline
\end{tabular}
\end{table*}

\begin{figure*}[c]
    \centering
    \includegraphics[height=27 cm]{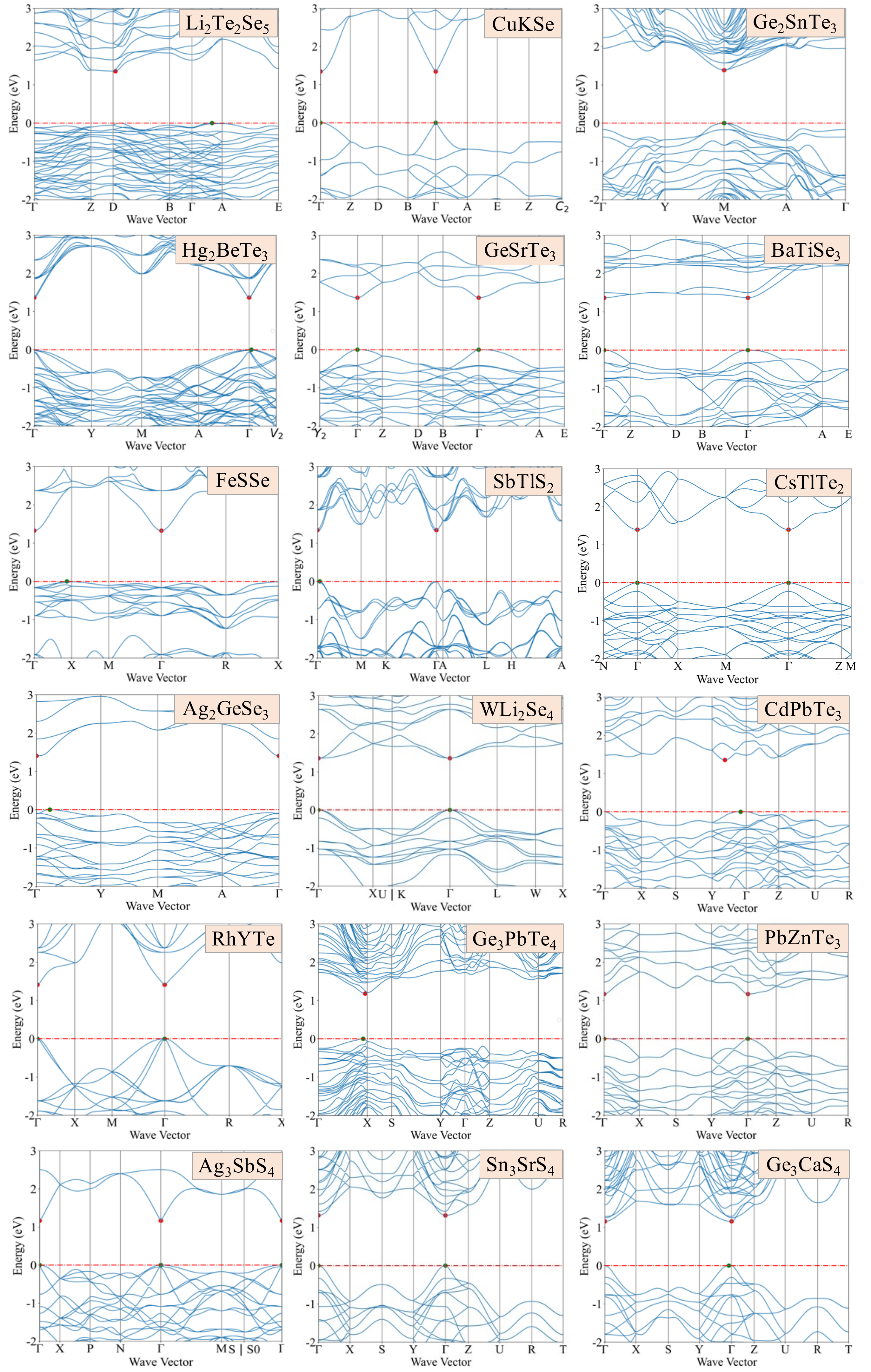}
    \label{figs5_1}
\end{figure*}

\begin{figure*}[c]
    \centering
    \includegraphics[height=27 cm]{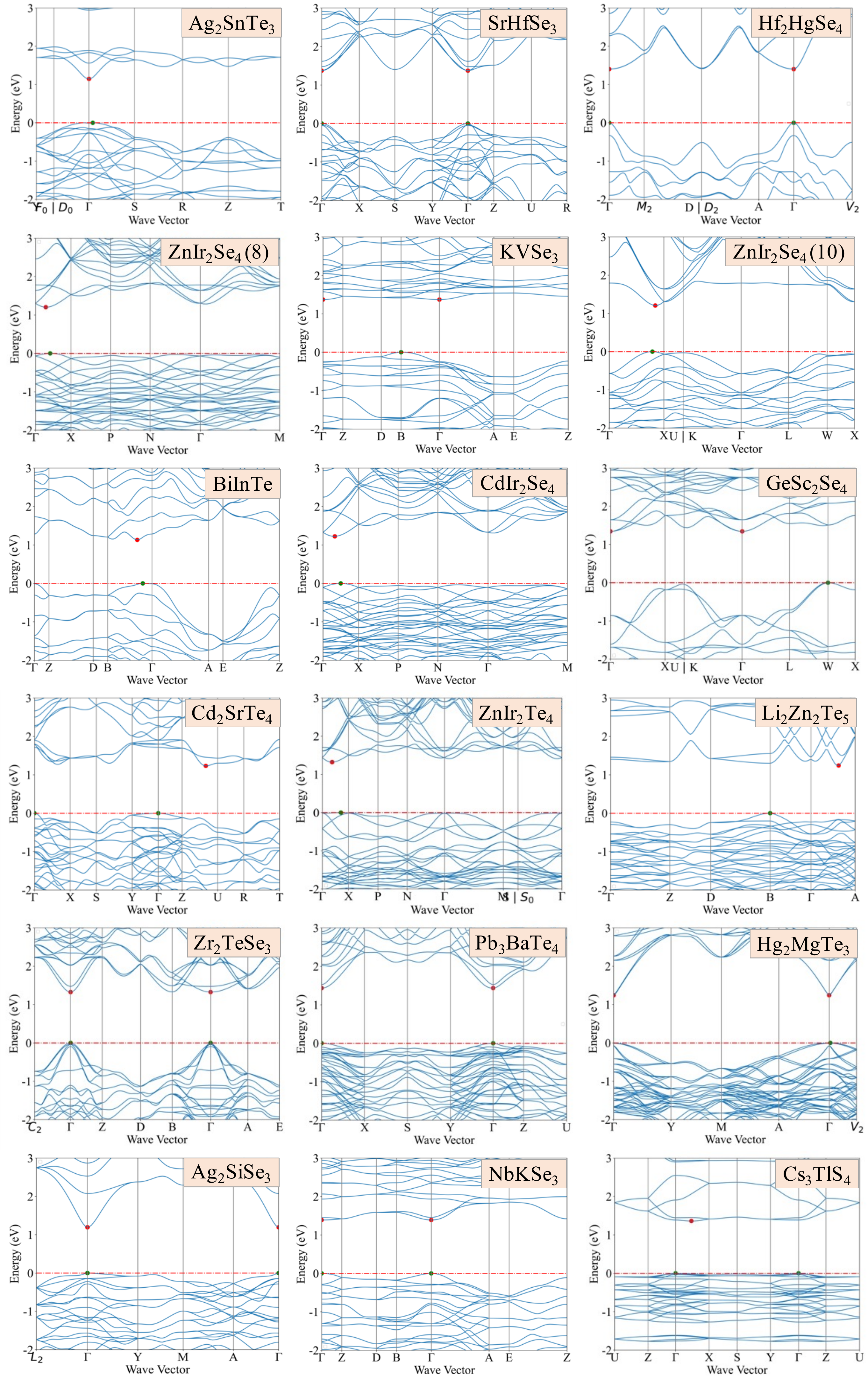}
    \label{figs5_2}
\end{figure*}

\begin{figure*}[c]
    \centering
    \includegraphics[height=27 cm]{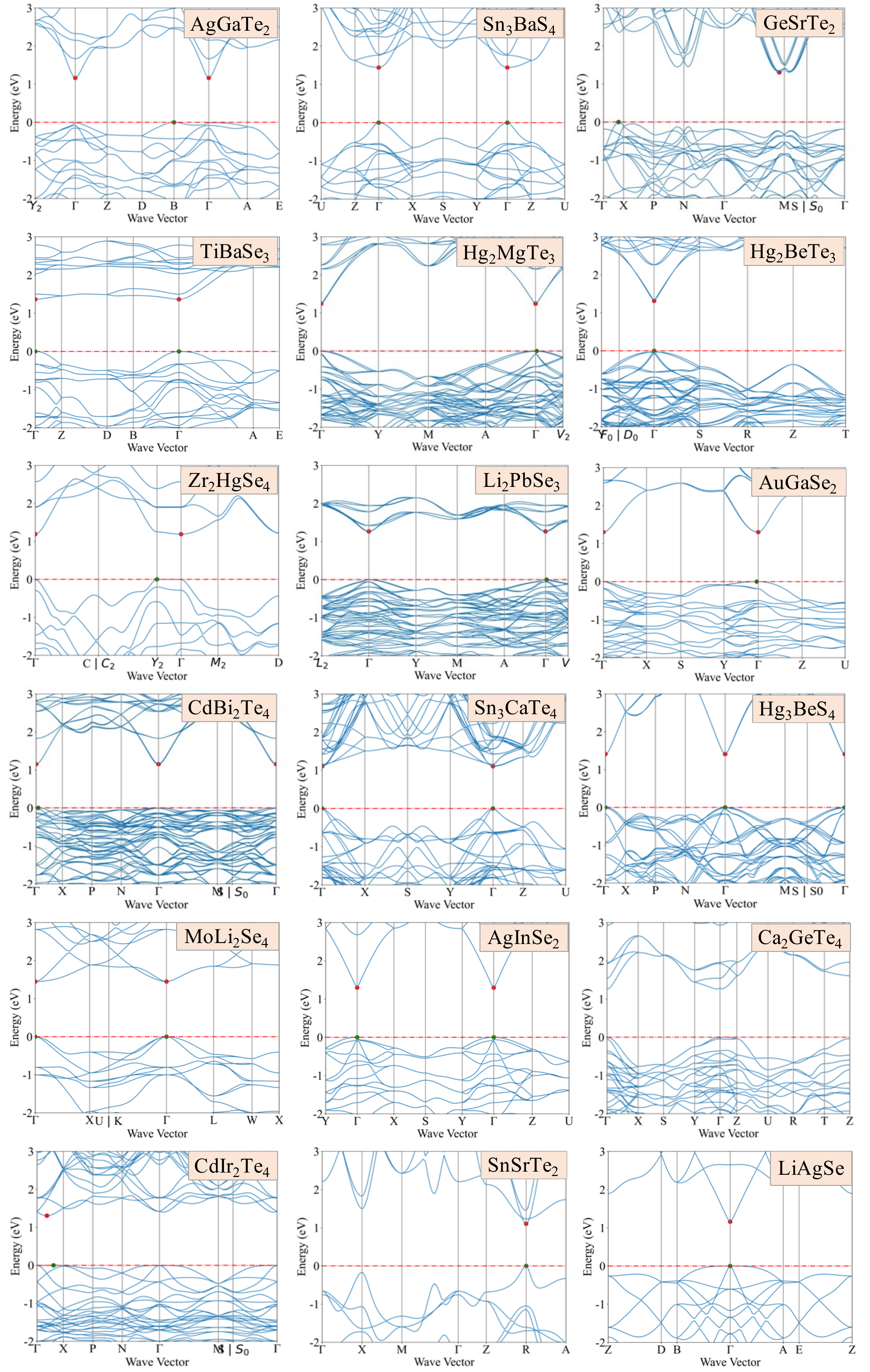}
    \label{figs5_3}
\end{figure*}

\begin{figure*}[c]
    \centering
    \includegraphics[height=18.5 cm]{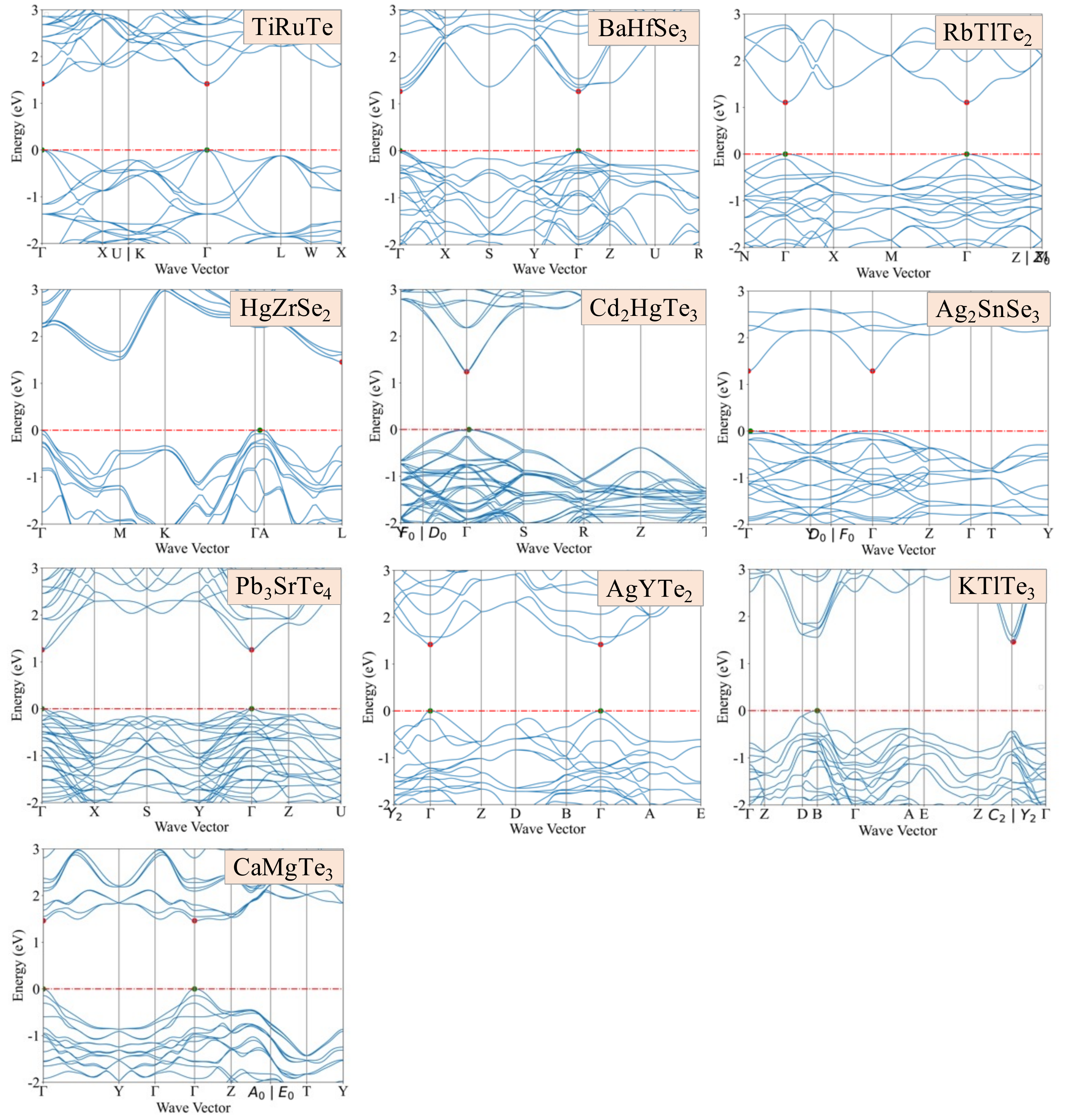}
    \label{figs5_4}
    \caption{Calculated band structures for 64 candidate compounds, using the scPBE0 functional}
\end{figure*}

\begin{figure*}[c]
    \centering
    \includegraphics[height=7 cm]{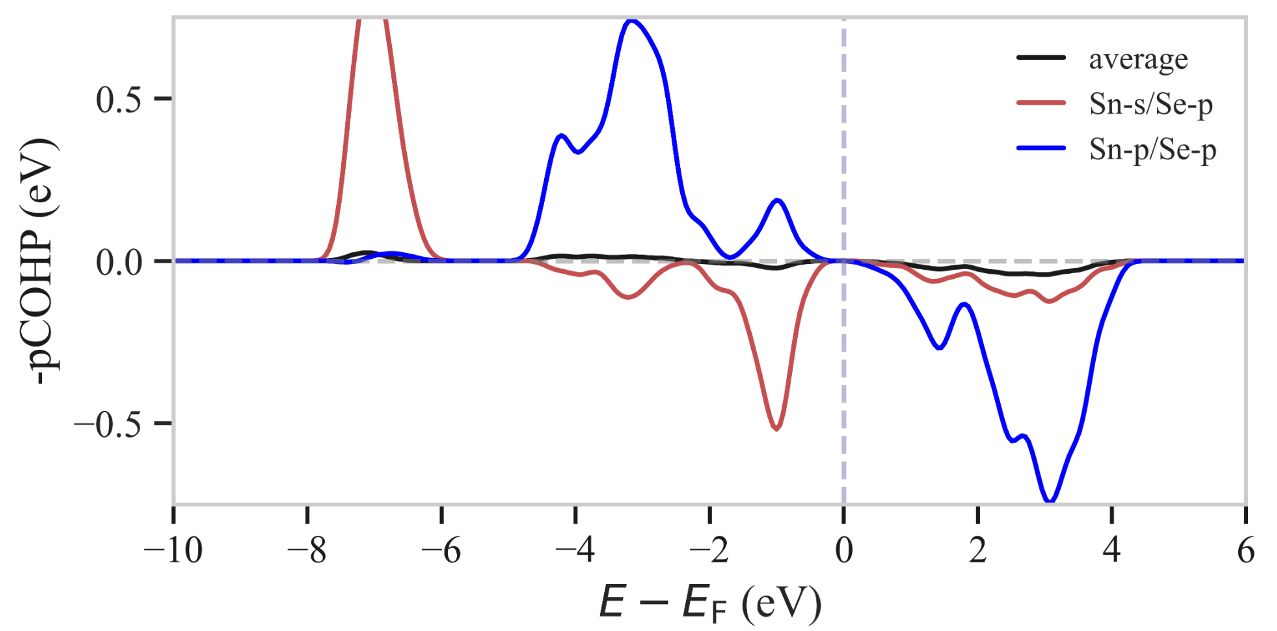}
    \label{figs6}
    \caption{Calculated projected crystal orbital Hamilton population of Sn$_3$BaSe$_4$. The Fermi energy is set to 0 eV.}
\end{figure*}

\bibliography{ref}